\documentclass[aps,prd,nopacs,nofootinbib,notitlepage]{revtex4-1}

\usepackage{amsfonts}
\usepackage{amsmath}
\usepackage{xcolor}
\usepackage{bm}
\usepackage{bbm}
\usepackage{graphicx}
\usepackage{fancybox}
\usepackage{comment}
\usepackage{multirow}
\usepackage{tipa}
\usepackage{longtable}
\definecolor{refs}{RGB}{245,156,74}
\usepackage[colorlinks=true,hyperfootnotes=true,citecolor=cyan]{hyperref}

\usepackage{enumitem}

\newcommand{\be}{\begin{equation}}
\newcommand{\ee}{\end{equation}}
\newcommand{\ba}{\begin{eqnarray}}
\newcommand{\ea}{\end{eqnarray}}
\newcommand{\bs}{\begin{subequations}}
\newcommand{\es}{\end{subequations}}

\newcommand{\mA}{\mathcal{A}}

\newcommand{\mS}{\mathcal{S}}

\newcommand{\hK}{\hat{K}}

\newcommand{\diff}{\textrm{d}}

\newcommand{\lp}{\left(}
\newcommand{\rp}{\right)}
\newcommand{\lb}{\left[}
\newcommand{\rb}{\right]}

\newcommand{\nn}{\nonumber}

\newcommand{\1}{1$^\text{st}$}
\newcommand{\2}{2$^\text{nd}$}
\newcommand{\3}{3$^\text{rd}$}

\newcommand{\so}{\mathfrak{so}}

\newcommand{\mpl}{M_{\text{Pl}}}

\begin{document}

\title{General Parallel Cosmology}

\author{D{\'e}bora Aguiar Gomes}
\email{debora.aguiar.gomes@ut.ee}
\address{Laboratory of Theoretical Physics, Institute of Physics, University of Tartu, W. Ostwaldi 1, 50411 Tartu, Estonia}

\author{Jose Beltr{\'a}n Jim{\'e}nez}
\email{jose.beltran@usal.es}
\affiliation{Departamento de Física Fundamental and IUFFyM, Universidad de Salamanca, E-37008 Salamanca, Spain}

\author{Tomi S. Koivisto}
\email{tomi.koivisto@ut.ee}
\address{Laboratory of Theoretical Physics, Institute of Physics, University of Tartu, W. Ostwaldi 1, 50411 Tartu, Estonia}
\address{National Institute of Chemical Physics and Biophysics, R\"avala pst. 10, 10143 Tallinn, Estonia}

\begin{abstract}

General (tele)parallel Relativity, G$_\parallel$R, is the relativistic completion of Einstein's theories of gravity. The focus of this article is the derivation of the homogeneous and isotropic solution in G$_\parallel$R. The first-principles derivation, based on a non-trivial realisation of the symmetry, supersedes and unifies previous constructions of Riemannian and teleparallel cosmologies, and establishes the uniqueness of the physical solution. The constitutive law and the form of the material and inertial source currents is presented in the tensor (Palatini) formalism and adapted to the cosmological background, which exhibits novel features absent in the previously studied, static solutions to the theory. The results are contrasted with those in incomplete theories, such as sitting at the three corners of the geometrical trinity which correspond to particular reference frames in G$_\parallel$R.   

\end{abstract}

\maketitle

\tableofcontents

\section{Introduction}

The standard concordance model of cosmology describes the large-scale observations successfully with just a few parameters \cite{Planck:2018vyg}. 
However, only a tentative more fundamental theory has been suggested to underpin the simplest, standard cosmological model \cite{Koivisto:2023epd}, whilst alternative, extended models abound \cite{CANTATA:2021ktz,Bahamonde:2021gfp}, further motivated by some significant anomalies in the experimental data \cite{Abdalla:2022yfr}.  
The foundational problems in cosmology concern energy: most of the energy of the universe is mysterious dark energy, most of the mass-energy is attributed to mysterious dark matter, and there is a disturbing mismatch of the energy scales in the present and in the early universe \cite{Abdalla:2022yfr,Perivolaropoulos:2021jda}. In attempts at quantum cosmology, one is faced with a quandary conjugated to energy, the notorious problem of time, see e.g. \cite{DiGioia:2019nti,Kiefer:2021zdq,Maniccia:2022iqa,Menendez-Pidal:2022tfm,Magueijo:2023poa} for some of the current discussions.

As it is well-known, energy can only be defined with respect to a reference \cite{https://doi.org/10.1002/andp.19163540702,Misner:1973prb,Szabados:2004xxa,Chen:2009zd}. According to a conventional wisdom, it reflects the equivalence principle that one cannot, in general, define the energy of a gravitational system. However, a complete theory of gravity should uniquely predict the observables such as energies and momenta. Whilst there may not exist an absolute, observer-independent separation of energetic interactions into gravitational vs. inertial effects, the physics should remain well-defined without the interactions and their associated energies somehow becoming meaningless or ambiguous once the metric geometry deviates from the flat-space standard. The resolution is that a given solution for the metric, flat or otherwise, provides that metric as the reference. More precisely, we had proposed that observables in gravity are resolved in the canonical frame defined by the vanishing of the point-wise localisable, covariant energy-momentum density of the metric field \cite{BeltranJimenez:2019bnx}. Subsequently, the proposal was proven to be unique in two respects.  
Firstly, it is consistent with the Noether theorems \cite{BeltranJimenez:2021kpj}. Secondly, it predicts the physical observables \cite{Gomes:2022vrc}.


The aim of this article is the derivation of the cosmological solution in the new, relativistic theory gravity. The principle of relativity posits that the laws of physics should assume the same form in all admissible reference frames. The literal incorporation of the principle into the theory of gravity is achieved via the extension of the covariance group consisting of the general coordinate transformations (Diff) by the symmetry group consisting of the general linear (GL) transformations. In a canonical frame, the conserved charges describe the observables, and in particular, the charge corresponding to time translations is the energy that is measured by probing the gravitational field. Therefore we can unambiguously pinpoint the physical properties such as energy, momenta, angular momenta, {\it etc} with the material source fields, with the metric field, or with the reference frame field. Given any metric, a physical observer is distinguished by the vanishing of the local energy-momentum associated with the reference frame.
At the conceptual level, this may be regarded as the realisation of the relativity of acceleration, an issue seldom recognised in the context of general relativity (GR) \cite{Sciama:1953zz,1992ASPC...30....1L,Oziewicz:2014qaa}.

At the technical level, the mathematical apparatus for the theory admits a geometric interpretation in terms of an affine structure which is flat, yet has non-trivial properties both with respect to a metric (non-metricity) and with respect to a coframe (torsion). Only a handful of studies has yet been carried into this general affinely-flat geometry \cite{BeltranJimenez:2019odq,Iosifidis:2018zwo,BeltranJimenez:2020sih,Hohmann:2021fpr,BeltranJimenez:2020sih,Heisenberg:2022mbo,Adak:2023ymc}. However, its two special limits accommodate the now often-studied reformulations of GR and their myriad modifications: the torsion-free case known as the symmetric teleparallel, and the metric-compatible case known as the metric teleparallel geometry, see e.g.  \cite{CANTATA:2021ktz,Bahamonde:2021gfp,Ferraro:2006jd,Jarv:2018bgs,Golovnev:2018wbh,Hohmann:2019nat,Raatikainen:2019qey,Jarv:2021ehj,Frusciante:2021sio,Pati:2021zew,Dimakis:2021gby,De:2022vfc,Albuquerque:2022eac,Ferreira:2022jcd,Lymperis:2022oyo,Esposito:2022omp,Gadbail:2023loj,Boehmer:2023knj,Dimakis:2023oje,Ferreira:2023kgv,Atayde:2023aoj,Paliathanasis:2023pqp,Shabani:2023xfn,Jarv:2023sbp} for their cosmological applications. The complementary descriptions of the gravitational interaction provided by those two alternatives in conjunction with the conventional one form the so called geometrical trinity of gravity \cite{BeltranJimenez:2019esp}, which has been discussed from various different perspectives in the recent literature \cite{BeltranJimenez:2017tkd,BeltranJimenez:2018vdo,Lu:2021wif,Capozziello:2022zzh,Nakayama:2022qbs,Koivisto:2022oyt,Erdmenger:2023hne,Wolf:2023rad,Bajardi:2023gkd,Garcia-Moreno:2023zdk,March:2023trv}. 
The version of GR in the generalised framework interpolating the base of the  geometrical trinity of gravity was called ``the general teleparallel equivalent of general relativity” \cite{BeltranJimenez:2019odq}, but since the relativistic theory we consider here is 
not an equivalent to GR\footnote{Also, G$_\parallel$R should be sharply distinguished from tenets commonly attributed to teleparallel gravity \cite{Bahamonde:2021gfp}. In particular, our universal definition of energy has essentially nothing to do with the ``local gravitational energy-momentum'' chased by some teleparallelists (see appendix \ref{literature}).}, we refer to it more briefly as G$_\parallel$R.


We present the formalism of the theory in Section \ref{para}, for generality deriving the equations for a generic (parity-even) quadratic action, which we then specialise to the case of G$_\parallel$R. Section \ref{parallelism} presents the construction of the cosmological reference matrix that determines the affine structure. Several distinct realisations of the isotropic and homogeneous symmetry are found to be possible, trivial and non-trivial. In Section \ref{cosmo} we apply the symmetric Ans{\"a}tze for the fields, to obtain the form of the constitutive law and the energy tensors in cosmology. A solution is found to the G$_\parallel$R field equations, and as expected, the solution is unique. Several cases of solutions for non-canonical frames are considered in more detail in the Appendices. In the concluding Section \ref{conclu} we discuss many possible directions for future investigations. 

\section{Paragravity}
\label{para}

Physical phenomena are interactions. We describe interactions in terms of fields.
Material sources can only be detected through the fields they excite. For instance, the charges of static or moving particles can only be detected by probing their electric or magnetic excitations, respectively. Therefore, the physical i.e. the measurable information is carried by field excitations. Physical observables are described as conserved charges, given by fluxes of field excitations on a closed surface (e.g. \cite{Penrose:1982wp,Barnich:2001jy,Donnelly:2016rvo,Donnelly:2016auv,Gomes:2022vrc,Ciambelli:2022vot}). This is the operational definition of an observable. We stress that it is fundamental and universal to all physical phenomena.


The structure of an interaction is determined by the constitutive law, which dictates the relation of the field excitation to the fundamental degrees of freedom in the theory. It is customary to formulate interactions in terms of gauge fields, and the excitations are then given by the momentum conjugates of the gauge fields\footnote{In standard Yang-Mills theories, the conjugates are the duals of the gauge field strengths. In a proposal for a more fundamental formulation, the excitations are (determined from) independent fields, and thus the constitutive law is a consequence of dynamics \cite{Gallagher:2022kvv}. Remarkably, this formulation resolves the long-standing problem of the Noether's \1 theorem in the standard model gauge theories \cite{Baker:2021hly}, which was not yet satisfactorily addressed in Ref. \cite{Gomes:2022vrc}.}. 
Furthermore, the gauge symmetry determines how the interaction fields couple to matter by e.g. using the reducibility parameters of the gauge symmetry. This procedure leads to field equations for the gauge fields that equate the divergence of their excitations and the conserved currents of the matter 
sector\footnote{These equations often imply that the observables can be formally re-expressed as volume integrals over matter sources. However, the material content in a volume does not represent the fundamental observable, which should be particularly clear in the topologically non-trivial cases of charges without (material) sources. Examples are monopoles in particle physics \cite{Shnir:2005vvi} and black holes in gravity \cite{Gomes:2022vrc}.}. In a \1 order formulation, this is nothing but the Hamilton equation that relates the sources with the variations of the conjugate momenta. 

In this Section we consider the structure of the gravitational interaction. 
In particular, we shall determine the constitutive law and the source currents for the covariant completion of the original formulation of GR \cite{https://doi.org/10.1002/andp.19163540702}. We thus introduce the metric
tensor $g_{\mu\nu}$, and consider sources given by the material energy tensor $T^\mu{}_\nu$. In a Lagrangian formulation, the latter is obtained as the variation $T_{\mu\nu} = g_{\mu\nu}L_{\text{matter}} - 2\delta L_{\text{matter}}/\delta g^{\mu\nu}$, and the dynamics is completely described by the fields in $L_{\text{matter}}$, the metric and the matter fields. However, 
we can extract observable predictions from the theory only once we have determined the reference frame of an observer. The frame is described by an element of GL,
represented as the 16-component matrix $\Lambda^\alpha{}_\beta(x)$. Thus, these elements can be regarded as fundamental objects, but it is important to stress that they {\it do not introduce any new degrees of freedom nor any new parameters} into the theory.  
The (equivalence classes of) auxiliary objects $\Lambda^\alpha{}_\beta(x)$ are uniquely fixed by the G$_\parallel$R principle, as shall be clarified in the following. 

It is conventional to discuss gravity in terms of geometry. In the language of Palatini formalism for spacetime geometry, the observer's frame determines an affine connection $\Gamma^\alpha{}_{\mu\beta}$ which by construction has no curvature $R^\alpha{}_{\beta\mu\nu}$, 
\be \label{constraint}
\Gamma^\alpha{}_{\mu\beta}=(\Lambda^{-1})^\alpha{}_\rho\partial_\mu\Lambda^\rho{}_\beta 
\quad \Rightarrow \quad R^\alpha{}_{\beta\mu\nu} = 2\partial_{[\mu} \Gamma^\alpha_{\phantom{\alpha}\nu]\beta} + 2\Gamma^\alpha_{\phantom{\alpha}[\mu\lvert\lambda\rvert}\Gamma^\lambda_{\phantom{\lambda}\nu]\beta} = 0\,.
\ee
This flat connection guides the inertial parallel transport in a spacetime given by $g_{\mu\nu}$. 
In parallel geometry, since the connection is curvature-free by construction, its only non-trivial geometrical properties are the torsion and the non-metricity given by 
\bs
\ba
T^\alpha{}_{\mu\nu} & = & 2\Gamma^\alpha{}_{[\mu\nu]}\,, \\ 
Q_\alpha{}^{\mu\nu} & = & -\nabla_\alpha g^{\mu\nu}\,, 
\ea
respectively. 
It will be convenient to introduce the notations for the traces of these tensors,
\be
T_\mu = T^\alpha{}_{\mu\alpha}\,, \quad 
Q_\alpha = g_{\mu\nu}Q_\alpha{}^{\mu\nu}\,, \quad
\tilde{Q}^\mu = Q_\alpha{}^{\mu\alpha}\,, \quad 
V^\mu = Q^\mu - \tilde{Q}^\mu + 2T^\mu\,. 
\ee
\es
Though the $\nabla_\alpha$ with the coefficients (\ref{constraint}) will be determined from a given metric in a canonical frame, it should not be confused with the Levi-Civita connection of the metric. 
We shall thus be working in a particularly simple special class of the generic Palatini geometry, which could be called the (general) parallel geometry. However, it is worth pointing out that the physics we discuss is independent of its possible geometrical depictions, and indeed there are equivalent formulations in terms of different fields in different geometrical frameworks, e.g. \cite{Koivisto:2018aip,Koivisto:2019ejt,Hohmann:2021fpr,Adak:2023ymc}.

\subsection{Energy tensors}
\label{energytensors}

Let us now consider a gravity theory described by the Lagrangian $L=L(g^{\mu\nu},Q_\alpha{}^{\mu\nu},T^\alpha{}_{\mu\nu})$ in the parallel geometry. It is convenient to introduce the non-metricity and torsion conjugates defined as
\begin{subequations}
\label{conjugates}
\ba
P^\alpha{}_{\mu\nu} & = &  \frac{\partial L}{\partial Q_\alpha{}^{\mu\nu}}\,, \\ \label{qconju}
S_\alpha{}^{\mu\nu} & = &  \frac{\partial L}{\partial T^\alpha{}_{\mu\nu}}\,.\label{tconju}
\ea
\end{subequations}
We can then write the equations of motion for the metric and the parallel connection, respectively, as
\bs
\label{eoms}
\ba
2\left({\nabla}_\alpha + T_\alpha + \frac{1}{2}Q_\alpha\right) {P}^\alpha{}_{\mu\nu} & = & g_{\mu\nu}L - 2\frac{\partial {L}}{\partial g^{\mu\nu}} + T_{\mu\nu}\,, \\
\left({\nabla}_\mu + T_\mu + \frac{1}{2}Q_\mu\right) \lp  S_\alpha{}^{\mu\nu} - P^{\mu\nu}{}_\alpha\rp & = & 0\,.
\ea
\es
We can now aim at realising the statement that sources are divergences of excitations with the gravitational excitations now considered as the conjugates of the metric and the parallel connection. To that end, it is convenient to introduce the metrical and the canonical energy tensors \cite{Gomes:2022vrc}
\bs
\ba
G_{\mu\nu} & = & g_{\mu\nu}L - 2\frac{\partial {L}}{\partial g^{\mu\nu}}
- 2Q_{\alpha\beta\mu} P^{\alpha\beta}{}_\nu\,, \label{Menergy} \\
t_{\mu\nu} & = & G_{\mu\nu} - T_{\mu\alpha\beta} S_{\nu}{}^{\alpha\beta}\,, \label{inertia}
\ea
\es
respectively. Then, using (\ref{eoms}), we can express the divergences of the conjugates (\ref{conjugates}) as
\bs
\ba
2\lp\nabla_\alpha + T_\alpha\rp\lp \sqrt{-g}P^{\alpha\mu}{}_\nu\rp & = &
\sqrt{-g}\lp {G}^\mu{}_\nu + {T}^\mu{}_\nu\rp\,, \label{eoms2a} \\
2\lp \nabla_\alpha + T_\alpha\rp\lp \sqrt{-g} {S}_\nu{}^{\alpha\mu}\rp 
- \sqrt{-g}T^\mu{}_{\alpha\beta} S_{\nu}{}^{\alpha\beta} & = & \sqrt{-g}\lp {t}^\mu{}_\nu 
+ {T}^\mu{}_\nu\rp\,. \label{eoms2b}
\ea
\es
From this last equation it follows, for an arbitrary vector $\xi^\mu$, that
\be \label{naive}
2\partial_\alpha\lp  \sqrt{-g} {S}_\nu{}^{\alpha\mu}\xi^\nu \rp = \sqrt{-g}\lp t^\mu{}_\nu + T^\mu{}_\nu \rp \xi^\nu + 2\sqrt{-g}S_\nu{}^{\alpha\mu}\nabla_\alpha \xi^\nu\,. 
\ee
Choosing any  vector $\xi^\mu$ such that the last term vanishes, this suggests the use of the Stokes theorem to convert the volume integral of the canonical and the material energy tensors into a surface integral
of the excitation tensor density. 
The suggested integral defines a charge with respect to an arbitrary $\xi^\mu$. The definition coincides with the expression deduced axiomatically from \1 principles \cite{Koivisto:2019ggr}. Furthermore, the exact same expression for the gravitational excitation tensor density, $2\sqrt{-g}S_\nu{}^{\alpha\mu}$, is the canonical Noether (super)potential whose integral is the canonical Noether charge \cite{BeltranJimenez:2021kpj}. The identification of the unique Noether charges as the observables of the theory is consistent with the implication of (\ref{naive}) that only the charges with respect to a Diff generated by a {\it constant} vector $\xi^\mu$ are conserved according to the naive coordinate-dependent interpretation.  

Note that $G_{\mu\nu}$ is not in general equivalent to the metric Einstein tensor familiar from the standard formulation of GR. We will see that the $G_{\mu\nu}$ reduces to the metric Einstein tensor in a zero energy frame (defined by $S_\alpha{}^{\mu\nu}=0$) in symmetric teleparallelism (more generally, when $T_{\mu\alpha\beta} S_{\nu}{}^{\alpha\beta}=0$). On the other hand, in a coincident frame (defined by $\Gamma^\alpha{}_{\mu\nu} = 0$) the $G_{\mu\nu}$ reduces to the so called Einstein pseudotensor. Thus, the metrical energy tensor $G_{\mu\nu}$ is the GL generalisation of both the metric Einstein tensor and the Einstein pseudotensor. 

In the original formulation of GR, the fundamental equation was written in a coordinate-dependent form. 
The minimal ``covariantisation'' of the original GR results in the coincident GR, an equivalent theory in symmetric teleparallel geometry \cite{Nester:1998mp,Adak:2005cd,BeltranJimenez:2017tkd,Koivisto:2018aip}. In contrast to the case of symmetric teleparallelism, in G$_\parallel$R the metrical energy tensor can now be non-vanishing $G^\mu{}_\nu \neq 0$ in a canonical frame defined by $t^\mu{}_\nu=0$. Then, the coincident gauge is not available because the anholonomy obstructs the elimination of a non-trivial $\Lambda^\mu{}_\nu$. 
In this concrete sense, G$_\parallel$R is not merely an equivalent reformulation of GR but its proper extension. 

\subsection{Quadratic theory}

We will now focus on the general parity preserving quadratic parallel gravity \cite{BeltranJimenez:2019odq}. This theory is constructed out of the quadratic non-metricity scalars defined as
\bs
\label{scalars}
\be
\label{Qscalars}
C_1  =  Q_{\alpha\mu\nu}Q^{\alpha\mu\nu}\,, \quad
C_2  =  Q_{\alpha\mu\nu}Q^{\mu\nu\alpha}\,, \quad
C_3  =  Q_\alpha Q^\alpha\,, \quad
C_4 = \tilde{Q}_\alpha\tilde{Q}^\alpha\,, \quad
C_5 = Q_\alpha \tilde{Q}^\alpha\,,
\ee
the quadratic torsion scalars
\be
\label{Tscalars}
A_1 =  T_{\alpha\mu\nu}T^{\alpha\mu\nu}\,, \quad
A_2 =  T_{\alpha\mu\nu}T^{\mu\alpha\nu}\,, \quad
A_3 = T_\alpha T^\alpha\,,
\ee
and the mixed scalars
\be
\label{Mscalars}
B_1 = T_{\alpha\mu\nu}Q^{\mu\nu\alpha}\,, \quad
B_2 = T_\alpha Q^\alpha\,, \quad
B_3 = T_\alpha \tilde{Q}^\alpha\,. 
\ee
\es
The quadratic parallel theory is then described by the following Lagrangian:
\be \label{quadratic}
L =  a_1 A_1 + a_2 A_2 + a_3 A_3 + b_1 B_1 + b_2 B_2 + b_3 B_3 +c_1 C_1+ c_2 C_2+c_3 C_3+c_4 C_4 + c_5 C_5\,, 
\ee
where $a_i,b_i,c_i$ are the parameters of the theory. We are most concerned with the particular case that reproduces the dynamics of GR. The dynamical equivalent can be straightforwardly obtained from the geometrical identity that relates the curvature scalar of the Levi-Civita connection $\mathcal{R}(g)$ with the torsion and non-metricity of a parallel geometry:
\bs
\ba
\mathcal{R}(g) & = & -\frac14 A_1 - \frac12 A_2 + A_3 - B_1 +  B_2 - B_3 -\frac14 C_1+ \frac12 C_2+\frac14 C_3 -\frac12 C_5 +\frac{1}{\sqrt{-g}}\partial_\mu\left(\sqrt{-g} V^\mu\right)\,,
\ea
\es
where $V^\mu  =  \tilde{Q}^\mu-Q^\mu-2T^\mu$. Using this relation, we can see that the Einstein-Hilbert action of GR can be obtained from our quadratic Lagrangian upon the choice of the following specific values of the parameters 
\be \label{gr}
c_1
= -\frac{c_2}{2}
= - c_3 
= \frac{c_5}{2} = -\frac{\mpl^2}{8}\,, \,\, 
c_4= 0\,, \quad\quad
a_1
= \frac{a_2}{2}
= -\frac{a_3}{4} =-\frac{\mpl^2}{8}\,, \quad\quad
b_1 
= -b_2 
= b_3 = -\frac{\mpl^2}{2}\,.  
\ee
With these parameters the two actions $I_{\rm GR}=-(\mpl^2/2)\int\diff^4 x\sqrt{-g}\mathcal{R}(g)$ and $I=\int\diff^4 x\sqrt{-g} L$ only differ by a boundary term and, hence, give rise to the equivalent field equations. For this choice of parameters, there is an enhanced gauge symmetry that allows to remove the parallel connection completely \cite{BeltranJimenez:2019odq}.
Nevertheless, for the sake of generality and completeness, we will derive the equations for the general 11-parameter theory. 

We first note that the quadratic Lagrangian (\ref{quadratic}) can be written in terms of the conjugates (\ref{conjugates}) as
\be \label{action}
L = \frac{1}{2}\Big( Q_\alpha{}^{\mu\nu}P^\alpha{}_{\mu\nu}
+ T^\alpha{}_{\mu\nu}S_\alpha{}^{\mu\nu}\Big)\,, 
\ee
where the explicit expressions of the conjugates are
\bs
\ba
P^\alpha{}_{\mu\nu} & = &  2 c_1 Q^\alpha{}_{\mu\nu} + 2c_2Q_{(\mu\nu)}{}^\alpha +  2c_3 Q^\alpha g_{\mu\nu} +  2c_4\delta^\alpha_{(\mu}\tilde{Q}_{\nu)}
+ c_5\lp \tilde{Q}^\alpha g_{\mu\nu} + \delta^\alpha_{(\mu}Q_{\nu)}\rp-b_1 T_{(\mu\nu)}{}^\alpha  + b_2T^\alpha g_{\mu\nu}
+ b_3 \delta^\alpha_{(\mu}T_{\nu)}\,, \quad \\
S_\alpha{}^{\mu\nu} & = & 2a_1T_\alpha{}^{\mu\nu} - 2a_2 T^{[\mu\nu]}{}_\alpha -2 a_3\delta^{[\mu}_\alpha T^{\nu]} + b_1 Q^{[\mu\nu]}{}_\alpha - b_2\delta^{[\mu}_\alpha Q^{\nu]} - b_3\delta^{[\mu}_\alpha\tilde{Q}^{\nu]}\,. 
\ea
\es
It is now tedious but straightforward to obtain
\ba
\frac{\partial L}{\partial g^{\mu\nu}} & = & 
c_1\lp Q_{\mu\alpha\beta}Q_\nu{}^{\alpha\beta} - 2Q_{\alpha\beta\mu}Q^{\alpha\beta}{}_\nu\rp 
- c_2Q_{\alpha\beta\mu}Q^{\beta\alpha}{}_\nu
 +  c_3\lp Q_\mu Q_\nu - 2Q^\alpha Q_{\alpha\mu\nu}\rp - c_4\tilde{Q}_\mu\tilde{Q}_\nu
- c_5\tilde{Q}^\alpha Q_{\alpha\mu\nu} \nn \\ 
& + & 
a_1\lp 2T_{\alpha\beta\mu}T^{\alpha\beta}{}_\nu - T_{\mu\alpha\beta}T_\nu{}^{\alpha\beta}\rp + a_2 T_{\alpha\beta\mu}T^{\beta\alpha}{}_\nu + a_3T_\mu T_\nu  
\nn \\
& - & b_1\lp T_{\alpha\beta(\mu}Q_{\nu)}{}^{\alpha\beta} + T_{(\mu\lvert \alpha\beta\rvert}Q^{\alpha\beta}{}_{\nu)}\rp
 +  b_2\lp T_{(\mu}Q_{\nu)} - T^\alpha Q_{\alpha\mu\nu}\rp \nn \\
 & = &  \frac{1}{2}Q_{(\nu}{}^{\alpha\beta}P_{\mu)\alpha\beta}-Q_{\alpha\beta(\mu}P^{\alpha\beta}{}_{\nu)} 
 - \frac{1}{2}T_{(\mu|\alpha\beta}S_{|\nu)}{}^{\alpha\beta}
 + T_{\alpha\beta(\nu}S^{\alpha\beta}{}_{\mu)}\,, \label{variation}
\ea
so that the metrical and canonical energy tensors are given by
\bs
\ba
G^\mu{}_\nu & = & \delta^\mu_\nu L - Q_\nu{}^{\alpha\beta}P^\mu{}_{\alpha\beta} + T^\mu{}_{\alpha\beta}S_\nu{}^{\alpha\beta} - 2T_{\alpha\beta\nu}S^{\alpha\beta\mu}\,,  \\
t^\mu{}_\nu & = & \delta^\mu_\nu L - Q_\nu{}^{\alpha\beta}P^\mu{}_{\alpha\beta}  - 2T_{\alpha\beta\nu}S^{\alpha\beta\mu}\,. \label{inertial1}
\ea
\es
Note that we do not need the explicit symmetrisation of the last line of (\ref{variation}). If we nevertheless symmetrise each of the 4 terms on that line, we can arrive at the decomposition of the energy tensors into their symmetric and antisymmetric components,
\bs
\ba
G_{\mu\nu} & = & g_{\mu\nu}L - Q_{(\mu}{}^{\alpha\beta}P_{\nu)\alpha\beta} + T_{(\mu}{}^{\alpha\beta}S_{\nu)\alpha\beta} - 2T_{\alpha\beta(\mu}S^{\alpha\beta}{}_{\nu)} - 2Q_{\alpha\beta[\mu}P^{\alpha\beta}{}_{\nu]}\,, \\
t_{\mu\nu} & = & g_{\mu\nu}L - Q_{(\mu}{}^{\alpha\beta}P_{\nu)\alpha\beta} - 2T_{\alpha\beta(\mu}S^{\alpha\beta}{}_{\nu)} - 2Q_{\alpha\beta[\mu}P^{\alpha\beta}{}_{\nu]} 
- T_{[\mu}{}^{\alpha\beta}S_{\nu]\alpha\beta}\,. \label{inertial2}
\ea
\es
An explicit expression for the canonical energy tensor is
\ba \label{einertial}
t^\mu{}_\nu & = & c_1\lp C_1 \delta^\mu_\nu -
2Q^{\mu\alpha\beta}Q_{\nu\alpha\beta}\rp 
+ c_2\lp C_2 \delta^\mu_\nu  - 2Q^{\alpha\beta\mu} Q_{\nu\alpha\beta}\rp \nn \\
& + & c_3\lp C_3 \delta^\mu_\nu - 2Q^\mu Q_\nu\rp + c_4\lp C_4\delta^\mu_\nu - 2\tilde{Q}_\alpha Q_\nu{}^{\mu\alpha}\rp 
 +  c_5\lp C_5\delta^\mu_\nu - Q_\alpha Q_\nu{} ^{\mu{}\alpha} - Q_\nu\tilde{Q}^\mu\rp \nn \\ & + & a_1\lp A_1\delta^\mu_\nu - 4 T_{\alpha\beta\nu}T^{\alpha\beta\mu}\rp + a_2\lp A_2\delta^\mu_\nu - 2T_{\alpha\beta\nu}T^{\beta\alpha\mu} + 2T_{\alpha\beta\nu}T^{\mu\alpha\beta}\rp 
 +  a_3\lp A_3\delta^\mu_\nu - 2T^\mu T_\nu + 2T^\alpha T^\mu{}_{\nu\alpha}\rp \nn \\ & + & b_1\lp B_1\delta^\mu_\nu + T_{\alpha\beta}{}^\mu Q_\nu{}^{\alpha\beta} 
- T_{\alpha\beta\nu}Q^{\beta\alpha\mu} + T_{\alpha\beta\nu} Q^{\mu\alpha\beta}\rp + b_2\lp B_2\delta^\mu_\nu + T^\mu{}_{\nu\alpha}Q^\alpha - T^\mu Q_\nu - T_\nu Q^\mu\rp \nn \\
& + & b_3\lp B_3\delta^\mu_\nu - T_\alpha Q_\nu{}^{\mu\alpha} + \tilde{Q}^\alpha T^\mu{}_{\nu\alpha} - \tilde{Q}^\mu T_\nu\rp\,.   
\ea
In the following we shall fix the parameters to those given in (\ref{gr}), and thus
focus on the case of G$_\parallel$R within the general quadratic parallel theory. 

\subsection{Summary of G$_\parallel$R}

The constitutive law that determines the theory (\ref{action})
is
\bs
\label{potentials}
\ba
S_\alpha{}^{\mu\nu}  & = & -\frac{\mpl^2}{2}\lb Q^{[\mu\nu]}{}_\alpha - \delta^{[\mu}_\alpha V^{\nu]}  + \frac{1}{2}T_\alpha{}^{\mu\nu} - T^{[\mu\nu]}{}_\alpha \rb\,,   \\
P^\alpha{}_{\mu\nu} & = & \frac{\mpl^2}{4}\lb -Q^\alpha{}_{\mu\nu} + 2Q_{(\mu\nu)}{}^\alpha - \delta^\alpha_{(\mu}\lp Q_{\nu)} + 2T_{\nu)}\rp - g_{\mu\nu}V^\alpha
+ 2T_{(\mu\nu)}{}^\alpha  \rb\,.
\ea
\es
The connection and the metric are to be solved from, respectively, the 2 sets of equations
\bs
\label{gprsum}
\ba
t^\mu{}_\nu & = &  \delta^\mu_\nu L - Q_\nu{}^{\alpha\beta}P^\mu{}_{\alpha\beta}  - 2T_{\alpha\beta\nu}S^{\alpha\beta\mu} = 0\,, \label{gprframe} \\
 T^\mu{}_\nu & = &
2\partial_\alpha\lp  \sqrt{-g} {S}_\nu{}^{\alpha\mu}\rp/\sqrt{-g} - 2S_\beta{}^{\alpha\mu}\Gamma^\beta{}_{\alpha\nu}\,, \label{efe} 
\ea
\es
Note that the system (\ref{gprsum}) is, despite appearances, both GL-invariant and Diff-covariant, since (\ref{efe}) is simply the rewriting of the GR field equation for the metric in a canonical frame determined by (\ref{gprframe}). To recapitulate the symmetries of the theory, the {\it frame transformation} is the GL parameterised by a matrix $\lambda^\mu{}_\nu$,
\bs
\label{frametrans}
\ba
\delta_\lambda g^{\mu\nu} & = & 0\,, \\
\delta_\lambda \Gamma^\alpha{}_{\mu\nu} & = & \nabla_\mu\lambda^\alpha{}_\nu\,, 
\ea
\es
and the {\it gauge transformation} is the Diff parameterised by a vector $\xi^\mu$ 
\bs
\ba
\mathcal{L}_\xi g^{\mu\nu} & = & 2\nabla^{(\mu}\xi^{\nu)} + \lp Q_\alpha{}^{\mu\nu} - 2T^{(\mu\nu)}{}_\alpha\rp\xi^\alpha\,, \\
\mathcal{L}_\xi \Gamma^\alpha{}_{\mu\nu} & = & \nabla_\mu\nabla_\nu\xi^\alpha + \nabla_\mu\lp T^\alpha{}_{\beta\nu}\xi^\beta\rp\,.  
\ea
\es
The GL freedom is fixed by (\ref{gprframe}). Then (\ref{efe}) provides the gauge-invariant description of the dynamics in a canonical frame. We shall now apply this to cosmology. 

\section{Homogeneous and isotropic parallelism}
\label{parallelism}

Cosmology is the most common application of teleparallel gravity models, and 
interesting considerations of cosmological symmetries in this context include \cite{Hohmann:2019fvf,Coley:2019zld,Hohmann:2020zre,Coley:2022qug}. In this Section, we will construct in a systematic way the homogeneous and isotropic teleparallel geometry in terms of the fundamental object $\Lambda^\alpha{}_\beta$, from which we can then derive the affine connection according to the point of departure (\ref{constraint}). The reader not interested in the construction of parallel geometry from the \1 principles is invited to skip this Section and move directly towards the cosmological application in the next Section \ref{cosmo}.

Our derivation will exploit the fact that the parallel connection can be written in terms of an element of the general linear group GL$(4,\mathbbm{R})$ as 
\be
\Gamma^\alpha{}_{\mu\beta}=(\Lambda^{-1})^\alpha{}_\rho\partial_\mu\Lambda^\rho{}_\beta\,,
\ee
so it features and invariance under global GL$(4,\mathbbm{R})$ transformations 
\be
\Lambda^\rho{}_\mu(x)\rightarrow A^\rho{}_\lambda\Lambda^\lambda{}_\mu(x)\,.
\ee
for a constant $A^\rho{}_\lambda\in \text{GL}(4,\mathbbm{R})$. In terms of this symmetry, the General Teleparallel Equivalent of GR corresponds to the particular quadratic theory where this global symmetry (that is present in any teleparallel theory constructed with the flat connection) becomes a local symmetry, i.e., the choice of parameters \eqref{gr} promotes the global symmetry to a local one (see e.g. \cite{BeltranJimenez:2019odq} for more details on this point). In the general parallel geometry, without assuming any particular theory, the realisation of the cosmological symmetries on $\Lambda^\alpha{}_\mu$ can be understood in terms of a symmetry breaking pattern where the original $\text{ISO}(3,1)\times \text{GL}(4,\mathbbm{R})$ group is broken down to some residual $\text{ISO}(3)$, $\text{SO}(4)$ or $\text{SO}(3,1)$  subgroup for the flat, closed and open cosmologies respectively. In standard cosmological scenarios, these symmetries are trivially realised on both the gravity and the matter sectors in the sense that the generators of the residual cosmological symmetry are given in terms of the generators of the original Poincar\'e group. However, the cosmological symmetries can also be non-trivially realised with interesting phenomenological consequences (see e.g. \cite{Maleknejad:2012fw,Endlich:2012pz,Nicolis:2015sra,Piazza:2017bsd,BeltranJimenez:2018ymu,Jimenez:2022fll,Aoki:2022ylc}). The non-trivial realisations consist in using some internal symmetries (global or gauge) so that, despite having background fields that break some of the spatial rotations and/or translations, the presence of the internal symmetries allows for the existence of unbroken generators that permit to realise the cosmological symmetries. For our parallel geometry, it will be the global $\text{GL}(4,\mathbbm{R})$ symmetry that will allow to realise the cosmological symmetries with  a background configuration for $\Lambda^\alpha{}_\mu$ that  breaks the spatial rotations and homogeneity and the cosmological group will be preserved by some diagonal combination of the corresponding subgroups of $\text{ISO}(3,1)$ and $\text{GL}(4,\mathbbm{R})$. There will be some differences between the three different symmetry groups of the flat, open and closed cosmologies that we will discuss in the following, but in all cases we will find that parallel geometry will contain two arbitrary functions of time associated to the connection. These functions correspond to a time-reparameterisation and a time-dependent dilation respectively. Let us then proceed to analysing each case separately.

\subsection{Spatially flat case}

The flat cosmology is characterised by a residual $\text{ISO}(3)$ group. It is obvious that homogeneity is trivially realised by a homogeneous $\Lambda^\alpha{}_\mu$, while rotations can be realised as a combination of a spatial rotation and a GL-transformation corresponding to the inverse rotation, which is possible because $\text{SO}(3)$ is a subgroup of  $\text{GL}(4,\mathbbm{R}$). The configuration that realises this symmetry breaking pattern in the simplest manner can be written as
\be
\hat{\Lambda}=
 \left(
\begin{array}{c|c}
\mu(t) &0    \\
  \hline
0  &\sigma(t)\mathbbm{1}  
\end{array}
\right)\,,
\label{eq:Trivalconfiguration}
\ee
with $\mu$ and $\sigma$ two arbitrary functions of time. We can perform a spatial rotation described by $R^i{}_j\in \text{SO}(3)$ together with an internal GL-transformation corresponding to a rotation in the 4-dimensional representation of $\text{ISO}(3)$, i.e., 
\be
\hat{A}=\left(
\begin{array}{c|c}
  1&0   \\
  \hline
0  &O^i{}_j  
\end{array}
\right)\,,
\ee
with $\hat{O}$ an orthogonal matrix. The inertial connection then changes under the simultaneous action of both transformations as follows:\footnote{As customary, we denote $(R^{-1})^i{}_j\equiv R_j{}^i$ so that $R^i{}_jR_k{}^j=\delta^i_k$.}
\ba
\Lambda^i{}_m(\vec{x})&\to& A^i{}_\alpha R_m{}^n \Lambda^\alpha{}_n(\hat{R}\cdot\vec{x})=\sigma O^i{}_k R_m{}^n\delta^k_n\,,
\ea
so we only need to choose $\hat{O}=\hat{R}$ to preserve the background configuration.\footnote{This realisation of the cosmological symmetry is analogous to the one that makes use of gauge fields as in e.g. \cite{Maleknejad:2012fw}.} This simply means that the change induced by a spatial rotation can be undone by the action of an internal transformation, thus leaving the configuration unchanged. The remaining components realise both homogeneity and isotropy in a trivial manner so we will spare the details.

The configuration \eqref{eq:Trivalconfiguration} is the simplest realisation of the residual $\text{ISO}(3)$ because homogeneity is trivially realised, hence, we will dub it the trivial realisation. However, there is another way of realising the residual Euclidean group for the parallel geometry that is more interesting because also homogeneity is realised in a non-trivial way. The key observation to understand the existence of an alternative realisation is to notice that the whole 3-dimensional Euclidean group admits a representation within $\text{GL}(4,\mathbbm{R})$ so spatial translations can also be non-trivially realised. The elements that do this job can be parameterised as
\bs
\be
\hat{\Lambda} =
\left(
\begin{array}{c|c}
\mu(t)  &-\sigma(t)\lambda\vec{x}   \\
  \hline
0  &\sigma(t)\mathbbm{1}  
\end{array}
\right)\,,
\label{Eq:FlatB}
\ee
with $\lambda$ a constant parameter that we introduce to keep track of the deviations with respect to the trivial realisation of the $\text{ISO}(3)$ symmetry that is then recovered for $\lambda=0$. In this expression, we must understand $\vec{x}$ as representing $x_m\equiv \delta_{mi}x^i$, i.e., the Euclidean dual of the position vector. The inverse of this matrix can be easily computed and is given by
\be
\hat{\Lambda}{}^{-1}=
\left(
\begin{array}{c|c}
  \mu^{-1}(t)&\mu^{-1}(t)\lambda\vec{x}   \\
  \hline
0  &\sigma^{-1}(t)\mathbbm{1}  
\end{array}
\right)\,.
\ee
\es
This configuration not only breaks spatial rotations, but now also spatial translations are broken due to the explicit dependence on $\vec{x}$. However, for this specific dependence (that in particular is linear), we can see that  a spatial translation $\vec{x}\to\vec{x}+\vec{x}_0$ can be compensated with a GL transformation that belongs to the ISO(3) representation within $\text{GL}(4,\mathbbm{R})$, while rotations will be realised in a similar manner to the trivial realisation \eqref{eq:Trivalconfiguration}. More explicitly, if we perform a spatial ISO(3) transformation parameterised by the constant vector $\vec{x}_0$ and the orthogonal matrix $\hat{R}$, we can undo the change by simultaneously performing an internal $\text{GL}(4,\mathbbm{R})$ transformation parameterised by
\be
\hat{A}=
\left(
\begin{array}{c|c}
  1&\vec{t}_0   \\
  \hline
0  &\hat{O} 
\end{array}
\right)\,,
\ee
with $\vec{t}_0$ a constant vector and $\hat{O}$ an orthogonal matrix that must be appropriately chosen. To obtain the relation between both transformations we will look at $\Lambda^0{}_i$, which is the only component that transforms non-trivially:
\ba
\Lambda^0{}_i=-\sigma\lambda x_i&\to& A^0{}_\beta R_i{}^j\Lambda^\beta{}_j(\hat{R}\cdot\vec{x}+\vec{x}_0)\,\nonumber\\
&=&R_i{}^j\left[-\sigma\lambda\Big(R_j{}^kx_k+x_{0,j}\Big)\right]+t_{0,m}R_i{}^j\sigma\delta^m{}_j\nonumber\\
&=&-\sigma\lambda x_i-\sigma R_i{}^j\Big[\lambda x_{0,j}-t_{0,j}\Big]\,.
\ea
Although the configuration does not remain invariant for arbitrary transformations, it is invariant under a diagonal combination upon the choice $\vec{t}_0=\lambda \vec{x}_{0}$. As anticipated, this corresponds to choosing a representation of translations within $\text{GL}(4,\mathbbm{R})$ that compensates the spatial translation. Regarding rotations, they can be seen to be realised just as in the trivial case so imposing invariance of $\Lambda^m{}_i$ fixes $\hat{O}=\hat{R}$. This is sufficient to prove the realisation of the residual $\text{ISO}(3)$ symmetry, but we can corroborate it by explicitly computing the connection. We find that the only non-vanishing components are
\bs
\be
\Gamma^0{}_{00}=\partial_0\log\mu,\quad \Gamma^i{}_{0j}=\partial_0\log\sigma\delta^i_j,\quad \Gamma^0{}_{ij}=-\frac{\sigma}{\mu}\lambda\delta_{ij}\,.
\label{Eq:GammaFlatB}
\ee
that satisfy the relation
\be
\Gamma^0{}_{00}-\frac13\Gamma^i{}_{0i}=-\partial_0\log\left(\delta^{ij}\Gamma^0{}_{ij}\right)\,.
\ee
\es
Since homogeneity is non-trivially realised, we will refer to this realisation as non-trivial.

A question that naturally arises at this point is whether this non-trivial realisation admits other non-equivalent representations. Interestingly, the answer is that it indeed does. This can be simply understood by noticing that the translations of the Euclidean group can be represented in two different manners in its four dimensional representation, namely: either with the first row or, alternatively, with the first column of the matrix. Thus, we could also parameterise the non-trivial realisation of homogeneity with the following elements of the General Linear group:
\bs
\be
\hat{\Lambda}=
\left(
\begin{array}{c|c}
\mu(t)  &0  \\
  \hline
\mu(t)\lambda\vec{x}   &\sigma(t)\mathbbm{1}  
\end{array}
\right)\,,
\label{Eq:FlatA}
\ee
with the inverse
\be
\hat{\Lambda}{}^{-1}=
\left(
\begin{array}{c|c}
\mu^{-1}(t)  &0  \\
  \hline
-\sigma^{-1}(t)\lambda\vec{x}   &\sigma^{-1}(t)\mathbbm{1}  
\end{array}
\right)\,.
\ee
\es
In this case we must understand that $\vec{x}$ represents $x^i$. At a more technical level, this possibility reflects the existence of dual representations that lead to inequivalent configurations. The relation is apparent if we compute the dual representation of \eqref{Eq:FlatB} that is defined as
\be
\Lambda^\star\equiv \big(\Lambda^{-1}\big)^T= \left(
\begin{array}{c|c}
\mu^{-1}(t)  & -\sigma^{-1}(t)\lambda\vec{x}\\
  \hline
 0   &\sigma^{-1}(t)\mathbbm{1}  
\end{array}
\right)\,,
\ee
which coincides with \eqref{Eq:FlatA} upon the identification $\mu^{-1}\to\mu$ and $\sigma^{-1}\to\sigma$. The residual ISO(3) cosmological symmetry is realised in a similar manner as in the previous case and it can be obtained as inherited by the duality relation. We can also corroborate that the ISO(3) symmetry is preserved by computing the connection, whose non-vanishing components are
\bs
\label{branchAflat}
\be
\Gamma^0{}_{00}=\partial_0\log\mu\,,\quad \Gamma^i{}_{0j}=\partial_0\log\sigma\delta^i_j\,,\quad \Gamma^i{}_{j0}=\frac{\mu}{\sigma}\lambda\delta^i_j\,.
\label{Eq:GammaFlatA}
\ee
and satisfy the relation
\be
\Gamma^0{}_{00}-\frac13\Gamma^i{}_{0i}=\partial_0\log\Gamma^i{}_{i0}\,. 
\ee
\es
We can explicitly check that this is the connection corresponding to the dual representation, since the connections for both representations relate as
\be
\Gamma^\star_\mu=(\Lambda^\star)^{-1}\partial_\mu\Lambda^\star=\Lambda^T\partial_\mu(\Lambda^{-1})^T=-(\Lambda^{-1}\partial_\mu\Lambda)^T=-\Gamma^T_\mu\, ,
\ee
which is indeed the relation between \eqref{Eq:GammaFlatA} and \eqref{Eq:GammaFlatB}. Let us notice that the dual connection is a connection in the dual space, but, in this particular case, the isomorphism between both spaces provided by the flat Euclidean metric is the trivial one (the identity) and, thus, the dual connection will also be a good connection for our purpose here. In the next subsection dealing with spatially curved cases where the metric is no longer the Euclidean one, this will no longer be possible, although, of course, dual connections can still be constructed.

This completes our construction of the spatially flat cosmologies that make a total of three realisations, namely: the trivial realisation \eqref{eq:Trivalconfiguration} and the two non-trivial realisations \eqref{Eq:FlatB} and \eqref{Eq:FlatA}. Let us notice that the trivial realisation is recovered by sending $\lambda\to0$ in the non-trivial ones. Let us now proceed to the spatially curved cosmologies.

\subsection{Spatially curved case}

Spatially curved cosmologies realise homogeneity by enlarging the isotropy group $\text{SO}(3)$ to either $\text{SO}(4)$ or $\text{SO}(3,1)$ for the closed and open universes respectively. Thus, homogeneity is characterised by  invariance under pseudo-translations given by rotations involving a fourth dimension and boosts. We can now use the fact that these two groups are subgroups of $\text{GL}(4,\mathbbm{R})$ to obtain non-trivial realisations of the cosmological symmetry by means of a combination of spatial transformations and internal ones. A parameterisation of the reference frame can be chosen as 
\be \label{BranchAref}
\hat{\Lambda} =
\left(
\begin{array}{c|c}
\mu\chi  &\frac{-k\sigma\ell\vec{x}}{\chi}   \\
  \hline
\mu \ell^{-1}\vec{x}  &\sigma\mathbbm{1}  
\end{array}
\right)\,,
\ee
with $\chi\equiv \sqrt{1-kr^2}$ and $k$ the curvature parameter. The constant $\ell$ determines a length scale, for which one could choose the conventional normalisation such that $k\ell^2 = \pm 1$.
In Cartesian coordinates, we can take the generators of spatial rotations and (pseudo-)translations as\footnote{In these expressions $\epsilon_{ijk}$ denotes the Levi-Civita symbol and indices are raised and lowered with the Euclidean metric.}
\bs
\be
J_i=\varepsilon_{ij}{}^kx^j\partial_k\,,\quad P_i=\chi\partial_i\,,
\ee
that satisfy the Lie algebra
\be
\left[J_i,J_j\right]=-\varepsilon_{ij}{}^kJ_k\,,\quad \left[P_i,P_j\right]=-k\varepsilon_{ij}{}^kJ_k\,,\quad \left[J_i,P_j\right]=-\varepsilon_{ij}{}^k P_k\,,
\ee
\es
that corresponds to the Lie algebras of SO(4) and SO(3,1) for $k>0$ and $k<0$ respectively (as well as recovering the Lie algebra of the 3-dimensional Euclidean group for $k=0$). The task is now to check that the action of these generators can be compensated with generators of the internal symmetry group $\text{GL}(4,\mathbbm{R})$. It is not difficult to convince oneself that this is indeed possible and the required internal generators are given by those of the corresponding $\text{SO}(4)$ and $\text{SO}(3,1)$ subgroups of $\text{GL}(4,\mathbbm{R})$ for $k>0$ and $k<0$ respectively.\footnote{In metric teleparallelism, the reference frame $\Lambda^\alpha{}_\mu$ determines the metric up to a Lorentz transformation, thus reducing the internal symmetry group from $\text{GL}(4,\mathbbm{R})$ to its $\text{SO}(3,1)$ component. In this case, $\text{SO}(4)$ is no longer a subgroup of the internal symmetry. This is the reason why complex tetrads arise to parameterise closed cosmologies in those theories \cite{Ferraro:2011zb,Capozziello:2018hly,Hohmann:2019nat,Hohmann:2020vcv,Bahamonde:2022ohm}. We can see this explicitly by imposing the vanishing of the non-metricity. We can do this from the expressions given in Table \ref{ParallelFrames}. Imposing $Q_1= Q_2=0$ leads to $\mu=\mu_0n$ and $\sigma=\sigma_0 a$ for some constants $\mu_0$ and $\sigma_0$. Using these results and further imposing $Q_3=0$, we obtain $\left(\frac{\mu_0}{\ell\sigma_0}\right)^2=-k$ which clearly shows the necessity of having a complex tetrad for closed cosmologies, while the open cosmologies admit a real tetrad.} Thus, spatial rotations are compensated with the internal generators of the corresponding $\text{SO}(3)$ subgroups, while the (pseudo-)translations for the open and closed cosmologies will be compensated with boosts and the three rotations involving the fourth axis respectively.  It will be instructive to see how the symmetries are realised explicitly. In this case, it will be more convenient to work with the infinitesimal transformations. Thus, let us consider a spatial (pseudo-)translation generated by $P_i$ together with an internal $\text{GL}(4,\mathbbm{R})$ transformation generated by
\be
T_i=
\left(
\begin{array}{c|c}
0  & k\ell\vec{e}_i   \\
  \hline
-\frac{1}{\ell}\vec{e}_i  &\hat{0}  
\end{array}
\right)\,,
\ee
with $\vec{e}_i$ a unit vector in the $i-$direction and the parameter $\ell$ has been introduced for convenience. These matrices generate boosts for $k<0$ and rotations for $k>0$ so they can indeed compensate the corresponding transformations for the non-flat cosmological symmetries generated by $P_i$. In order to corroborate this claim, let us first notice that the combined action of $P_i$ and $T_i$ changes $\Lambda^\alpha{}_\mu$ as
\ba
\delta_i\Lambda^\alpha{}_\mu&\equiv& \mathcal{L}_{P_i}\Lambda^\alpha{}_\mu+(T_i)^\alpha{}_{\beta} \Lambda^\beta{}_\mu\nonumber\\
&=&\chi\partial_i\Lambda^\alpha{}_\mu-\frac{k}{\chi}x_m\delta_\mu^m\Lambda^\alpha{}_i+(T_i)^\alpha{}_{\beta} \Lambda^\beta{}_\mu\,.
\ea
We can check that this variation indeed vanishes: 
\bs
\ba
\delta_i\Lambda^0{}_0&=&\chi\partial_i\Lambda^0{}_0+(T_i)^0{}_0\Lambda^0{}_0=\chi\mu\partial_i\chi+k\delta_{im}\mu x^m=0\,,\\
\delta_i\Lambda^m{}_j&=&\chi\partial_i\Lambda^m{}_j-\frac{k}{\chi}x_j\Lambda^m{}_i+(T_i)^m{}_{0}\Lambda^0{}_j=-\frac{k}{\chi}x_j\sigma\delta^m_i-\delta_i^m\left(-\frac{k\sigma x_j}{\chi}\right)=0\,,\\
\delta_i\Lambda^j{}_0&=&\chi\partial_i\Lambda^j{}_0+(T_i)^j{}_{0}\Lambda^0{}_0=\ell^{-1}\lp\chi\partial_i(\mu x^j)-\delta_i^j\mu\chi\rp=0\,,\\
\delta_i\Lambda^0{}_j&=&\chi\partial_i\Lambda^0{}_j-\frac{k}{\chi}x_j\Lambda^0{}_i+(T_i)^0{}_{m}\Lambda^m{}_j=\ell\lb\chi\partial_i\left(-\frac{k\sigma x_j}{\chi}\right)+\frac{k^2\sigma}{\chi^2}x_ix_j+k\sigma\delta_{ij}\rb\nonumber\\
&=&-k\ell\sigma\left[\delta_{ij}+\left(\chi\partial_i\frac{1}{\chi}-\frac{k}{\chi^2}x_i\right)x_j\right]+k\ell\sigma\delta_{ij}=0\,.
\ea
\es
In the above computations it is useful to notice that $\chi\partial_i\chi=-k x_i$. We thus see that $\hat{\Lambda}$ is indeed invariant under the combined action of $P_i$ and $T_i$, i.e., homogeneity is achieved as a linear combination of the internal and spacetime transformations. The invariance under rotations is analogous to the flat case and can also be straightforwardly checked at the infinitesimal level so we will save the details to the reader. 

The inverse is now given by
\be
\hat{\Lambda}{}^{-1}=
\left(
\begin{array}{c|c}
\mu^{-1}\chi  &\mu^{-1}k\ell\vec{x}   \\
  \hline
-\frac{\chi}{\ell\sigma} \vec{x}  &\frac{1}{\sigma}(\delta^i_j-kx^ix_j)
\end{array}
\right)\,,
\label{Eq:InvLambdaNF}
\ee
and the non-vanishing components of the connection can be written as
\bs
\label{branchA}
\ba
\Gamma^0{}_{00}=\partial_0\log\mu\,,\quad \Gamma^i{}_{0j}=\partial_0\log\sigma\delta^i_j\,,\quad\Gamma^i{}_{j0}=\frac{\mu}{\ell\sigma}\delta^i_j\,, \quad
\Gamma^0{}_{ij}=-k\ell\frac{\sigma}{\mu}\gamma_{ij},\quad\Gamma^i{}_{jk}=kx^i\gamma_{jk}\,,
\ea
with 
\be
\gamma_{jk}=\delta_{jk}+\frac{kx_jx_k}{\chi^2}\,,
\label{eq:gammadef}
\ee
the metric of the maximally symmetric 3-space. The fact that we can write the connection in terms of this metric shows that we have indeed obtained a connection with the spatially curved cosmological symmetries, as can be checked by an explicit calculation. Moreover, let us notice that the purely spatial connection $\Gamma^i{}_{jk}$ is nothing but the Levi-Civita connection of the metric $\gamma_{ij}$. We can also check that the connection components satisfy the relation
\be
\frac19\gamma^{ij}\Gamma^0{}_{ij}\Gamma^m{}_{m0}=-k\,.
\ee
\es
From the reference frame \eqref{BranchAref} we can recover the flat case by taking appropriate limits. For this, it is convenient to note that a rescaling of the constant $\ell$ corresponds to
performing a global spatial dilation. 
The three spatially flat configurations can then be obtained by taking the following limits:
\begin{itemize}
\item Trivial flat configuration. The configuration given in \eqref{eq:Trivalconfiguration} can be obtained by taking $k\to0$, $\ell\to\infty$ with $k\ell\to 0$.
\item Non-trivial configuration I. The configuration \eqref{Eq:FlatB} is recovered by taking the limit $k\to0$, $\ell\to\infty$ while keeping $k\ell\equiv\lambda$ fixed.
\item Non-trivial configuration II. The second non-trivial configuration given in \eqref{Eq:FlatA} is recovered by simply taking $k\to0$ while keeping $\ell$ finite. We then identify $\lambda=\ell^{-1}$.
\end{itemize}
At the level of the connection components, it is easy to check that they are related by taking the same limits, thus showing the full relation between the different flat configurations. We show explicitly how to obtain the spatially flat configurations from the curved ones in Table \ref{ParallelFrames} for all the relevant quantities. The above limits to obtain the spatially flat cosmologies from the curved ones can be understood in terms of In\"on\"u-Wigner group contractions where the spatial translations after the contraction are realised either by the pure spatial translations $P_i$ or the two independent linear combinations of the external and internal ones involving $P_i$ and $T_i$. We will present a more thorough discussion of this construction elsewhere and now we will turn our attention to another inequivalent realisation that exists for the closed cosmologies.

So far, we have implicitly assumed that parity is preserved. If we allow for parity violation, one can show that the closed cosmologies admit another inequivalent parameterisation. The idea relies on the fact that the closed cosmologies correspond to SO(4), whose Lie algebra decomposes into (is isomorphic to) the direct sum of two copies of the SO(3) Lie algebra $\so(4)\cong\so(3)\oplus\so(3)$, so one copy generates isotropy and the other copy generates homogeneity. One could then wonder if it would be possible to use only one copy of the internal $\so(3)$ to compensate for the full $\so(4)$ of the external transformations, i.e., if one single internal $\so(3)$ can be used to realise both homogeneity and isotropy in combination with the external ones. In order to show that this is indeed possible, let us consider the following reference frame matrix
\be
\hat{\Lambda} =
\left(
\begin{array}{c|c}
\mu(t)  &0   \\
  \hline
0  &\sigma(t)\left(\chi\delta^i_j+\frac{1}{\chi}\ell^{-2} x^ix_j-k\ell\varepsilon^i{}_{jk}x^k\right)
\end{array}
\right)\,,
\label{eq:exceptionalframe}
\ee
and let us now consider a combined transformation generated by $\ell P_i$ and an internal rotation $(A_i)^m{}_n= \varepsilon_{i\,\,\,\,n}^{\,\,m}$. The only components that change are $\Lambda^m{}_n$ and they do so as
\be
\delta_i\Lambda^m{}_n=\left(k\ell^2-1\right)\frac{k\ell^2\sigma}{\chi}\,,
\ee
so we see that the configuration \eqref{eq:exceptionalframe} remains invariant provided $k\ell^2=1$, i.e., for the closed cosmologies, while the open ones do not admit such a configuration. The inverse of \eqref{eq:exceptionalframe} is given by
\be
\hat{\Lambda}{}^{-1}=
\left(
\begin{array}{c|c}
\mu^{-1}  &0   \\
  \hline
0  &\sigma^{-1}\left(\chi\delta^i_j+\frac{1}{\ell}\varepsilon^i{}_{jk} x^k\right)
\end{array}
\right)\,.
\ee
The non-vanishing components of the connection can then be easily computed and can be written as
\be
\Gamma^0{}_{00}=\partial_0\log\mu\,,\quad\Gamma^i{}_{0j}=\partial_0\log\sigma\delta^ i_j\,,\quad\Gamma^i{}_{jk}= k x^i\gamma_{jk}+\ell^{-1}\chi\varepsilon^{imn}\gamma_{mj}\gamma_{nk}\,.
\ee
We can express $\Gamma^i{}_{jk}$ in a more suggestive form by using that $\det\gamma=\chi^{-2}$ and introducing the volume form of the maximally symmetric space $\epsilon_{ijk}=\sqrt{\det\gamma}\varepsilon_{ijk}$ so $\epsilon^{ijk}=\chi\varepsilon^{ijk}$ and we have
\be
\Gamma^i{}_{jk}= k x^i \gamma_{jk}+\ell^{-1}\epsilon^i{}_{jk}.
\ee
This shows that this configuration introduces a totally antisymmetric constant torsion for the 3-space that was not present in the regular configurations. Furthermore, this configuration recovers the trivial configuration of the spatially flat cosmologies in the limit $k\to0$, $\ell\to\infty$ keeping the constraint $k\ell^2=1$ (i.e., $k\ell\to0$). Since we will consider parity preserving theories and configurations, this particular realisation of the cosmological symmetries will not be relevant for us and, as we shall show below, this configuration does not admit a canonical frame.\footnote{It may be important to notice that parity is among the GL-transformations that are not simply connected with the identity, but it is a symmetry of the parallel connection so, even a parity violating frame configuration might respect parity for the relevant physical quantities.} Let us emphasise once more that this realisation only exists for closed cosmologies because homogeneity is realised either with usual translations or boosts in the flat and open cosmologies respectively so we cannot play the same game (at least with real representations). We will thus refer to this configuration as exceptional.

The five inequivalent realisations of the cosmological symmetries unveiled here reproduce the five cosmological branches obtained in \cite{Hohmann:2020zre} and \cite{Heisenberg:2022mbo} from a more direct approach. There, the authors first work out the form of the connection respecting the cosmological symmetries and then solve the equations imposed by the flatness condition
(see appendix \ref{alternative} for more details). Our approach here instead has started from a flat connection directly in terms of the reference frame $\Lambda^\alpha{}_\mu$ and, then, we have exploited the existence of an internal global symmetry to impose the cosmological symmetries that are realised as combinations of these with the external Poincar\'e group. In both cases, the resulting connection in all the branches eventually depend on two functions of time and we have been able to give a physical meaning of those functions by providing a direct relation between those two functions and the fundamental object $\Lambda^\alpha{}_\mu$. It remains to show, in the next Section, that the two free functions are uniquely fixed by the cosmological metric in a canonical frame.

\section{Paracosmology}
\label{cosmo}

We now proceed to the next step in the main task of this work, i.e., the derivation of the cosmological solution to the parallel gravity theory. 
First, in \ref{ansatz} we state homogeneous and isotropic Ans\"atze for all the fields and the relevant geometrical objects constructed from them. To help comparison with the existing literature on teleparallel cosmology, we formulate the Ans\"atze in terms of the connection coefficients derived from the fundamental matrix $\hat{\Lambda}$.  
In \ref{gpc} we analyse the field equations in the physical branch of solutions in a generic frame. 

\subsection{The homogeneous and isotropic Ansatz}
\label{ansatz}
\begin{figure}[ht!]
\centering
\includegraphics[scale=0.7]{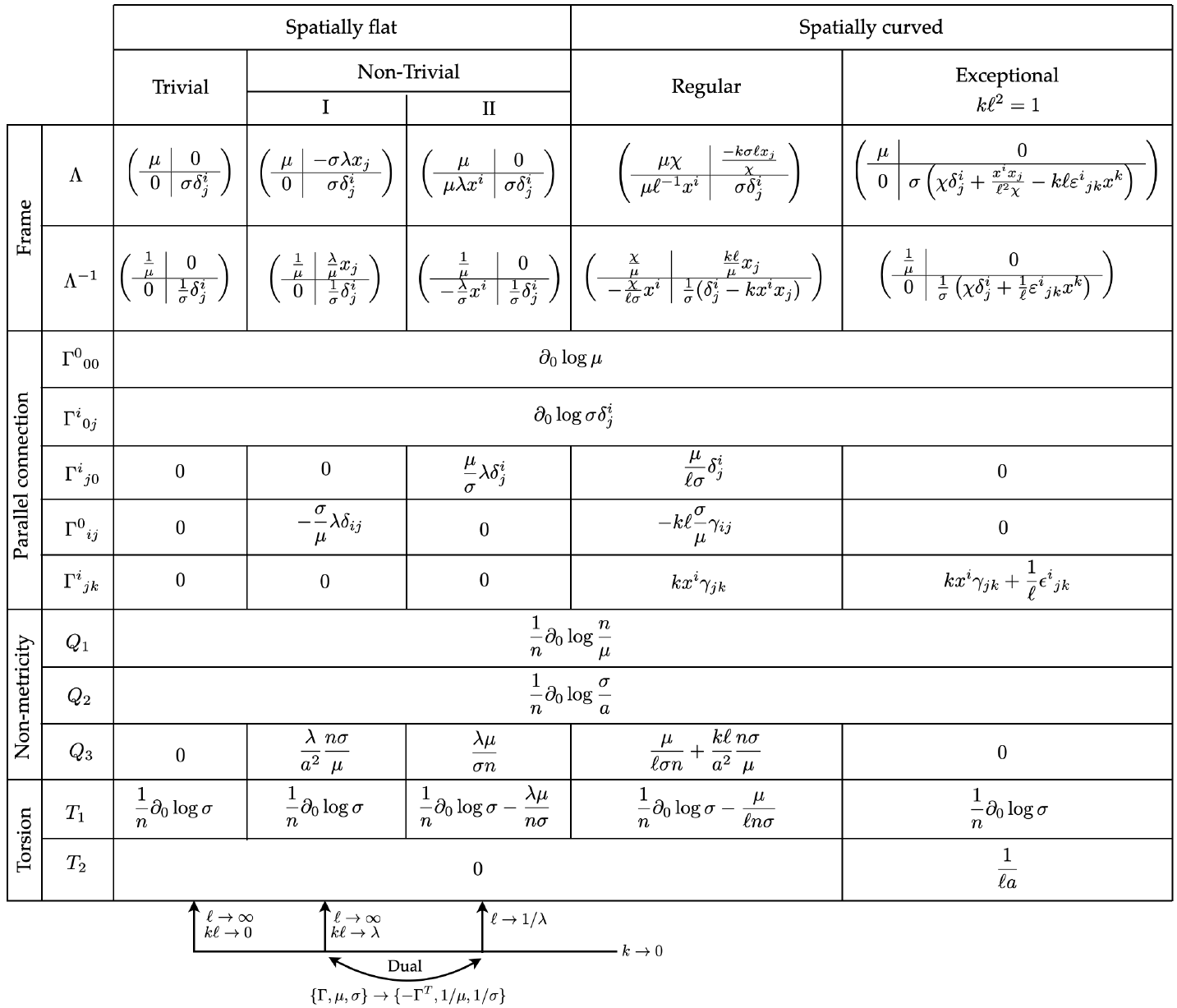}
\caption{This Table summarizes the cosmological parallel frames and the corresponding parallel connection components. We also give the torsion and non-metricity functions as defined in \eqref{eq:QTfunctions}. We also provide the limits that permit to go from the spatially curved cosmologies to the three spatially flat ones.}
\label{ParallelFrames}
\end{figure}

For the generic cosmological scenario, the spatial sections are assumed to be isotropic and homogeneous and, thus, are maximally symmetric so they correspond to constant curvature metric spaces up to a time-dependent conformal factor $a(t)$. This scale factor together with the lapse function $n(t)$ that determines the proper time conform the two independent functions that specify the cosmological Friedmann-Lema\^itre(-Robertson-Walker) described by the line element
\be
\diff s^2=-n^2(t)\diff t^2+a^2(t)\gamma_{ij}\diff x^i\diff x^j,
\ee
with $\gamma_{ij}$ the metric of maximally symmetric spaces given in \eqref{eq:gammadef}. It is convenient to employ the 1+3 decomposition of the metric with respect to the 4-velocity $u^\mu$ of a comoving observer that foliates the spacetime. We can then introduce the orthogonal projector
\be \label{metric}
h_{\mu\nu}=g_{\mu\nu} +u_\mu u_\nu \,,
\ee
that describes the geometry of the spatial hypersurfaces orthogonal to the 4-velocity. For the comoving observers we of course have $h_{ij}=a^2(t)\gamma_{ij}$. It is also convenient to define the associated expansion rates 
\be
N \equiv u^\mu\partial_\mu \log{n}\,, \quad H \equiv u^\mu\partial_\mu\log{a}\,. 
\ee
The physical quantities associated to the parallel connection, i.e. the torsion and the non-metricity, can be parameterised in the 1+3 decomposition as follows:
\bs
\label{eq:QTfunctions}
\ba
T^\alpha{}_{\mu\nu} & = & 2T_1 h^\alpha{}_{[\mu}u_{\nu]}
+ 2T_2 \epsilon^\alpha{}_{\mu\nu}\,, \label{ctorsion} \\
Q_\alpha{}^{\mu\nu} & = & 2Q_1 u_\alpha u^\mu u^\nu + 2Q_2 u_\alpha h^{\mu\nu}
+ 2Q_3 h_\alpha{}^{(\mu}u^{\nu)} \label{cnonmetricity} \,, 
\ea
\es
where the cosmological and torsion functions $T_a$ and $Q_a$ are not all independent and they are given in terms of the two free functions of the parallel connection derived in Sec. \ref{parallelism}. The explicit relation depends on the specific configuration under consideration and they are given in Table \ref{ParallelFrames}. Let us notice that the functions $Q_1$ and $Q_2$ are in turn the same for all the cosmological configurations, so the non-metricity only depends on the specific configuration through $Q_3$. As for the torsion functions, $T_2$ is non-vanishing only for the exceptional curved configuration. As we have argued before, this branch is neglected in the present work by imposing parity and we will also show below that selecting a canonical frame also requires having $T_2=0$, so the torsion will only have one free function given by $T_1$.

We have now all the necessary ingredients to proceed with the computation of the relevant quantities for the parallel cosmological scenario.

\renewcommand{\figurename}{Table}

\subsection{General parallel cosmology}
\label{gpc}

The cosmological Lagrangian for G$_\parallel$R can be written in terms of the non-metricity and torsion functions introduced above as
\bs
\label{isolag}
\ba 
\sqrt{-g}L & = & -\frac{3 \mpl^2}{2}na^3\left[ Q_1Q_3 + Q_2\Big( 2Q_2 - Q_3 - 4T_1\Big) + 2\Big( Q_3T_1 + T_1^2 - T_2^2\Big)\right] 
\,. 
\ea
We can express this Lagrangian in terms of the frame functions $\mu$ and $\sigma$. We will write it for the spatially curved regular configuration, from which all the spatially flat configurations can be obtained, and it is given by
\be
\sqrt{-g} L=3\mpl^2n a^3\left(\frac{k}{a^2}-H^2\right)+\frac32\mpl^2\partial_0\left(  \frac{a^3\mu}{\ell n\sigma}-\frac{k\ell a n\sigma}{\mu}\right).
\label{eq:LCosm}
\ee
\es
In this expression, we explicitly see how the connection only contributes a total derivative, as expected from the pure gauge character of the reference frame, and the pure metrical sector is the usual mini-superspace Lagrangian of GR. Furthermore, by taking the limits for the three spatially flat configurations we obtain that they can be distinguished by this boundary term since the trivial configuration ($\ell\to\infty$ and $k\ell\to0$) has a trivial boundary contribution, while the two non-trivial configurations, with ($\ell\to\infty$, $k\ell\to\lambda$) and ($\ell\to\lambda^{-1}$, $k\ell\to0$) for I and II respectively, have non-trivial boundary contributions given by each of the two boundary terms in \eqref{eq:LCosm}.

Let us now look at the constitutive laws. We can write it in terms of the non-metricity and torsion functions as
\bs
\ba
S_\alpha{}^{\mu\nu}  & = &  \mpl^2\Big(2T_1- 2Q_2 + Q_3 \Big) h_\alpha{}^{[\mu}u^{\nu]} + \frac{1}{2}\mpl^2T_2\epsilon_\alpha{}^{\mu\nu}  \,, \\
P^\alpha{}_{\mu\nu} & = & \frac{\mpl^2}{4}\Big[
3 Q_3u^\alpha u_\mu u_\nu + \big( 4Q_2-Q_3-4T_1\big) u^\alpha h_{\mu\nu} + 2\big( Q_1-Q_2+ 2T_1\big)h^\alpha{}_{(\mu}u_{\nu)}\Big] \,. 
\ea
\es
The metrical and the inertial energy tensors are given by
\bs
\ba
t^\mu{}_\nu & = & - 3\mpl^2\Big[ \lp Q_2-T_1\rp{}^2 + T_2^2\Big] u^\mu u_\nu - \mpl^2\Big[ Q_1Q_3 + \lp Q_2-T_1\rp\lp 3Q_2-Q_3-T_1\rp + T_2^2\Big]  h^\mu{}_\nu 
 \,,\label{eq:tmn} \\ 
G^\mu{}_\nu & = & t^\mu{}_\nu + \mpl^2\Big( 2Q_2 T_1 - Q_3T_1 - 2T_1^2 + 2T_2^2\Big) h^\mu{}_\nu  \,,
\ea
\es
and thus may differ only by an effective pressure.

We can now impose the condition $t^\mu{}_\nu=0$ to obtain the cosmological canonical frames. In principle, this is an extra condition we impose so, in general, it will not be possible to reach such a canonical frame. In fact, for an arbitrary quadratic theory this will not be possible because we only have a global symmetry for the frames. For the theory with parameters \eqref{gr}, we have a better chance because the symmetry is local and we can then choose an arbitrary local frame.\footnote{ This is due to the fact that the fields $\Lambda^\alpha{}_\mu$ only enter as total derivatives in the action so that they are pure gauge. This is one of the remarkable and distinguishing properties of GR.} However, it is not guaranteed that such a local freedom will suffice to achieve a canonical frame. In the following, we will show that this is indeed possible. From the expression of $t^\mu{}_\nu$ in \eqref{eq:tmn}, we can see that the condition of having a canonical frame requires the following three conditions:
\be
Q_2=T_1,\qquad T_2=0,\qquad Q_1Q_3=0.
\label{eq:condCF}
\ee
The condition $T_2=0$ excludes the curved exceptional configuration, so we will disregard it from now on, although we had already dismissed it from the requirements of parity invariance, so it is consistent to obtain that this configuration does not admit a physical branch of solutions. Let us notice that this result already shows the failure of metric parallelism because therein $Q_a=0$, in which case the torsion should identically vanish. On the other hand, in symmetric teleparallelism $T_a=0$, which forces $Q_2=0$ and the remaining condition imposes that there can only be one non-trivial function given by either $Q_1$ or $Q_3$. However, this possibility does not exist either as we will show below. Thus, we find that the existence of the cosmological canonical frames requires the entire parallel structure and, in particular, we see that the full $\text{GL}(4,\mathbbm{R})$ group is needed. 

For the canonical frames, the constitutive laws simplify substantially and are given by
\bs
\ba
S_\alpha{}^{\mu\nu}  & = &  \mpl^2Q_3 h_\alpha{}^{[\mu}u^{\nu]} \ \,, \\
P^\alpha{}_{\mu\nu} & = & \frac{\mpl^2}{4}\Big[
3 Q_3u^\alpha u_\mu u_\nu  -Q_3 u^\alpha h_{\mu\nu} + 2\lp Q_1-Q_2\rp h^\alpha{}_{(\mu}u_{\nu)}\Big] \,. 
\ea
\es
while the metrical energy tensor simplifies to
\be
G^\mu{}_\nu  =  t^\mu{}_\nu - \mpl^2Q_2Q_3 h^\mu{}_\nu  \,.
\ee
In particular, this expression shows that only when $Q_1=0$ (so we can have $Q_3\neq0$) can there be pressure differentiating the metrical and the canonical energy tensors.

In order to find whether there exists a cosmological canonical frame, we will consider directly the spatially curved cosmologies. We can then analyse the spatially flat cosmologies by taking the appropriate limits. We will first notice that the condition $T_2=0$ in \eqref{eq:condCF} simply excludes the exceptional configuration so we can disregard this configuration. The condition $Q_2=T_1$ results in the following relation
\be
H=\frac{\mu}{\ell n \sigma}\,.
\label{eq:Hcanonicalframe}
\ee
We can plug this relation in the expression for $Q_3$ to obtain
\be
Q_3=H\left(1+\frac{k}{a^2H^2}\right)\,.
\ee
Thus, the remaining condition $Q_1Q_3=0$ in \eqref{eq:condCF}, that in principle admits two branches with $Q_1=0$ and $Q_3=0$, actually only admits the branch $Q_1=0$ for having a cosmological evolution, since the branch $Q_3=0$ would only be possible in a curvature dominated universe. Thus, we further require $Q_1=0$. Since $Q_1=\frac{1}{n}\partial_0\log\frac{n}{\mu}$, we find that the frame function $\mu$ must be proportional to the lapse function, i.e., $\mu=\mu_0 n$ for some constant $\mu_0$. We then obtain that the canonical frame corresponds to the regular curved configuration with 
\be
\mu=\mu_0n,\quad \sigma=\frac{\mu_0}{\ell H}\,. 
\ee
This concludes our prove that it is indeed possible to construct a cosmological canonical frame for curved cosmologies. This frame is completely determined by the metric according to the above relations, i.e., the FLRW metric determines the G$_\parallel$R frame matrix to be
\be \label{frame}
\hat{\Lambda} =
\mu_0\left(
\begin{array}{c|c}
n\chi  & -\frac{k}{\chi H} x_j   \\
  \hline
n \ell^{-1}x^i  & (\ell H)^{-1}\delta^i_j
\end{array}
\right)\,.
\ee
Obviously, the constant $\mu_0$ can be removed by a global GL transformation so it does not play any physical role. At this point, we can show that this canonical frame does not exist for symmetric parallelism that further requires the additional constraint $T_1=0$. In this case, the equation $Q_2=T_1$ would result instead in $\partial_0\log a=0$, that imposes a constant scale factor and, consequently, there would not be any cosmological evolution.

From the result obtained for the curved cosmologies, we can now analyse the spatially flat cosmologies  by taking the corresponding limits as follows:
\begin{itemize}
\item Trivial configuration: $k\to0$, $\ell\to\infty$, $k\ell\to0$. When taking this limit, we see from \eqref{eq:Hcanonicalframe} that\footnote{We assume that both $\mu$ and $\sigma$ remain finite and non-vanishing in order to have a non-singular connection.} $H\to0$ so the scale factor must be constant and there is no cosmological evolution. Thus, this configuration does not admit a cosmological canonical frame. This could be concluded directly from the fact that the condition $Q_2=T_1$ for the trivial configuration reads $\partial_0\log a=0$ which, again, imposes a constant scale factor.

\item Non-trivial configuration I: $k\to0$, $\ell\to\infty$, $k\ell\to\lambda$. This configuration leads to the same problem as the trivial configuration so no canonical frame is possible either. This is easy to understand because it has the same functions $Q_2$ and $T_1$ as the preceding trivial configuration, so it will also require the absence of cosmological evolution.

\item Non-trivial configuration II: $k\to0$, $\ell\to\lambda^{-1}$. In this case, the limit is regular and there is no restriction on the cosmological evolution, so this configuration is the only spatially flat cosmology that admits a canonical frame. 
\end{itemize}

The fact that only the non-trivial configuration II admits a cosmological canonical frame shows the non-triviality for the existence of such a frame.

We can now go back to the constitutive laws and express them in the canonical frame
\bs
\ba
S_\alpha{}^{\mu\nu} & = & \mpl^2 H\lp 1 + \frac{k}{a^2H^2} \rp h_\alpha{}^{[\mu}u^{\nu]}\,, \\
P^\alpha{}_{\mu\nu} & = & \frac{\mpl^2}{4}\lb H\lp 1 + \frac{k}{a^2H^2}\rp\big(3 u_\mu u_\nu -  h_{\mu\nu}\big)u^\alpha - \frac{2}{n}\partial_0\log(aH)h^\alpha{}_{(\mu}u_{\nu)}\rb\,.
\ea
\es
These expressions adopt a more suggestive form if we introduce the usual curvature and slow-roll parameters defined as
\be
\Omega_k\equiv-\frac{k}{a^2H^2},\quad \epsilon\equiv-\frac{1}{nH}\partial_0\log H=-\frac{H'}{H^2}.
\ee
With these definitions the constitutive laws read
\bs
\label{potentials2}
\ba
S_\alpha{}^{\mu\nu} & = & \mpl^2 H\lp 1-\Omega_k\rp h_\alpha{}^{[\mu}u^{\nu]}\,, \\
P^\alpha{}_{\mu\nu} & = & \frac{\mpl^2 }{4} H\Big[\big( 1-\Omega_k\big) \lp 3u_\mu u_\nu - h_{\mu\nu}\rp u^\alpha - 2\lp 1 - \epsilon\rp h^\alpha{}_{(\mu}u_{\nu)}\Big]\,.
\ea
\es

Let us now have a look at the field equation. If we introduce the short-hand notation for the superpotential
\be
S_\alpha{}^{\mu\nu} = \mS h_\alpha{}^{[\mu}u^{\nu]}
\quad \text{where} \quad \mS = \mpl^2 H(1-\Omega_k)\, 
\ee
the G$_\parallel$R field equation then takes the form
\be
T^\mu{}_\nu = 3H \mS u^\mu u_\nu
- \lb {\mS}' + \left(\log\frac{a^3}{\sigma}\right)' \mS \rb h^\mu{}_\nu\,.
\ee
Finally, the metric field is associated with the energy tensor
\be \label{metricpressure}
G^\mu{}_\nu = \frac{2}{3}L h^\mu{}_\nu\,, \quad
\text{where} \quad L = \frac{ 3\mpl^2}{2}H^2\lp 1 - \epsilon\rp\lp 1-\Omega_k\rp\,.
\ee
This describes effective pressure, which is positive in an accelerating, and negative in a decelerating universe. 

\section{Conclusion}
\label{conclu}

In this article, the cosmological reference matrix $\hat{\Lambda}$ was constructed 
by nontrivial realisation of the isotropic and homogeneous symmetry. 
We have stated the constitutive law of G$_\parallel$R at (\ref{potentials}) and adapted it to cosmology at (\ref{potentials2}). The generic form of the energy tensors was given at (\ref{gprsum}), and this was found to imply that the metric field exerts an effective pressure (\ref{metricpressure}) in a cosmological canonical frame. The uniqueness of the canonical frame was established (up to the Diff, which in the isotropic and homogeneous case reduce to time reparameterisations), excluding 2 non-canonical branches of solutions in the spatially flat, and 1 non-canonical branch in the spatially curved case. The reference frames and their relations are summarised in Table \ref{ParallelFrames}.

From a static perspective, cosmological background observable is simply the expansion, quantified by the gauge-invariant Hubble rate $H$. The momentum charge in a given volume is $H$ times the area of the enclosing surface in Planck units. 
The possible non-vanishing energy charge must have an entirely topological origin, as it does in the case of a black hole \cite{Gomes:2022vrc}. A distinctive new feature of the cosmological solution with respect to the black hole solution is the presence of torsion which generates dissipation terms in the conservation equations. However, there are always 4 canonically given conserved currents, generated by the $\xi^\mu$ identified with the 4 columns of the $\Lambda^{-1}$. This yields just the expected energy integral over $\rho$. 
The full clarification of the energy of the Universe from a global perspective and its thermodynamic interpretation call for topological considerations which were beyond the scope of this article.

The equivalent description of dynamics is possible in alternative reference frames. Different reference frames can be advantageous for different purposes. For instance, it could make the computation of the dynamics of binary systems technically more feasible than the usual method of post-Newtonian perturbative expansion which has been developed only for special backgrounds. Another example is the problem of backreaction,
where the fully covariant treatment of the averaging problem requires a notion of the 
variation of the metric field, which can be appropriately defined only by referring
to a metric-independent covariant derivative. 

Thus, besides disclosing the physics in canonical frames, the variety of non-canonical frames available in G$_\parallel$R presents a potentially useful toolbox for tackling calculations in gravity. An example which might be relevant in the case that the manifold has a boundary, is the frame which makes the Lagrangian vanish at the boundary, $L\overset{\text{b}}{=}0$. This frame realises a natural smoothness principle for the action integral\footnote{In the Lorentz gauge theory of spacetime and gravitation \cite{Zlosnik:2018qvg}, $L\overset{\text{b}}{=}0$ is the prediction rather than a preferable choice of frame. The natural boundary is the “pregeometric" ground state $g_{\mu\nu}=0$ (not $g_{\mu\nu} = \eta_{\mu\nu}$) \cite{Koivisto:2023epd}.} 
which renders the on-shell action stationary with respect to arbitrary variations \cite{Koivisto:2022oyt}. 
\begin{itemize}
\item Smooth-boundary frame: $L \overset{\text{b}}{=}0$. This is only a partial frame fix in the canonical branch of solutions, whereas in the non-canonical Branch 1a and 2a there is no smooth-boundary frame.  
\end{itemize}
Some alternative reference frames mentioned in this article are as follows.
\begin{itemize}
\item Canonical frame: $t^\mu{}_\nu=0$. The Noether charges are the observable energy and momenta \cite{BeltranJimenez:2021kpj}.
\item Symmetric teleparallel frame: $T^\alpha{}_{\mu\nu}=0.$ This frame is not unique, but can be fixed into a canonical one iff the metric energy tensor can be eliminated by an integrable GL transformation \cite{Koivisto:2018aip}. This is not possible in generic cosmology. 
\item Metric teleparallel frame: $Q_\alpha{}^{\mu\nu}=0$. In the cosmological background, this fixes a non-canonical frame wherein the fictitious anti-energy is radiation-like.
\item Coincident frame: $\Gamma^\alpha{}_{\mu\nu}=0$ \cite{BeltranJimenez:2022azb,Bahamonde:2022zgj,Junior:2023qaq}. This is like the previous case, but with stiff anti-energy. 
\item Energy-free frame: $S_\alpha{}^{\mu\nu}=0$. Observers in this class of frames would measure no energy nor momenta. The similar result has been arrived at from different prescriptions in the literature, e.g. \cite{rosen,Xulu:2002ix,Nester:2008xd,vargas,deLaRica:2009kk,Ulhoa:2010wv,Lapiedra:2010av,Poplawski:2013mra,Johri:1995gh,Abedi:2022mvp}, often motivated by that vanishing total energy is in line with ideas of the Universe as a quantum fluctuation out of “nothing", e.g.  \cite{tryon,Vilenkin:1982de,Koivisto:2023epd}. 
\end{itemize}
Finally, let us comment on 2 examples of yet different possible formulations of the dynamics, which are not strictly frames according to our definition.
\begin{itemize}
\item Material (pseudo)frame: $t^\mu{}_\nu = 2S_\alpha{}^{\mu\beta}\Gamma^\alpha{}_{\beta\nu}$. In this (pseudo)frame, the charges are determined solely by the material energy tensor\footnote{This prescription conforms with some arguments of T. Levi-Civita, F. Klein, H. A. Lorentz etc \cite{earman1993attraction,Cooperstock1992-COOELI,Duerr:2019nar,Nikolic:2014kga,Wu:2016tzh,Aoki:2022gez}.}. This is not a proper equivalence class of frames, since the criterion is coordinate-dependent. 
\item  Hypothetical constant-rank frame in modified gravity. Though this article was devoted to the G$_\parallel$R cosmology, some of the derivations might be relevant also to teleparallel modified gravity models \cite{CANTATA:2021ktz,Bahamonde:2021gfp}. Such models could describe interesting cosmologies but their theoretical consistency of remains dubious since the degrees of freedom are generically ill-defined (e.g. \cite{Cheng:1988zg,Chen:2014qtl,BeltranJimenez:2020fvy,Blixt:2021anz,BeltranJimenez:2021auj,Barker:2022kdk}). It is an open question whether this could be fixed by a “properly parallelised'' prescription for the reference frame \cite{Koivisto:2018loq}, in view of that in modified gravity models such prescriptions change the dynamics rather than just the frame of reference.   
\end{itemize}

The conjecture that for any solution of GR there exists a unique (modulo Diff) canonical frame in G$_\parallel$R is not strictly proven. It will be interesting to generalise the isotropic and homogeneous G$_\parallel$R solution to anisotropic cosmologies as well as to proceed to the perturbative treatment of structure formation since there the physics is described in terms of more non-trivial observables (charges) and field force lines (excitations). 

\begin{acknowledgments}
This work was supported by the Estonian Research Council grant PRG356 “Gauge gravity: unification, extensions and phenomenology”. J.B.J. was supported by the Projects PGC2018-096038-B-100 and PID2021-122938NB-I00 funded by the Spanish “Ministerio de Ciencia e Innovaci\'on". 
\end{acknowledgments}

\appendix 

\section{An alternative parameterisation}
\label{alternative}

The frame matrix is the fundamental object from which we deduce the parallel affine connection. This is a foundational point of departure with respect to the previous literature on teleparallel cosmology,  wherein the starting point has been an isotropic and homogeneous Ansatz for the affine connection, which is then constrained to be flat. It might be useful to adapt (some of) our derivations also to the existing conventions in the literature.

In the following we will use the parameterisation and notations used by Hohmann \cite{Hohmann:2021ast} of the generic cosmological affine connection in terms of 5 time-dependent functions $K_i$, using the 1+3 decomposition which could be quoted as\footnote{The parameterisation of the general affine connection in terms of five functions of time has been previously used in e.g. \cite{Minkevich:1998cv,Iosifidis:2020gth}.}
\ba
\Gamma^\alpha{}_{\mu\nu}  & \approx & \frac{K_1}{n} u^\alpha u_\mu u_\nu  + \frac{n K_2}{a^2} u^\alpha h_{\mu\nu} - \frac{K_3}{n}  h^\alpha{}_\mu u_\nu- \frac{K_4}{n} u_\mu h^\alpha{}_\nu  
+ \frac{K_5}{a} \epsilon^{\alpha}{}_{\mu\nu} \nn \\
& \equiv & \hK_1 u^\alpha u_\mu u_\nu  + \hK_2 u^\alpha h_{\mu\nu} - \hK_3 u_\nu h^\alpha{}_\mu - \hK_4 u_\mu h^\alpha{}_\nu  
+ \hK_5 \epsilon^{\alpha}{}_{\mu\nu} \,, \label{connection}
\ea
where $\epsilon_{\alpha\mu\nu} \equiv u^\beta\epsilon_{\beta\alpha\mu\nu}$, and we
introduced the rescaled functions $\hK_i$ which will be convenient. The $\approx$ sign indicates that the apparently covariant parameterisation (of the form used in \cite{Iosifidis:2020gth}) of the non-tensorial object $\Gamma^\alpha{}_{\mu\nu}$ might actually be legitimate only in some coordinate system(s). Hohmann had considered the parameterisation adapted to spherical symmetry in the polar coordinates in \cite{Hohmann:2019fvf}, and the coordinate expressions (meriting the $=$ sign) can be found in the original references \cite{Hohmann:2019fvf,Hohmann:2021ast}. By the same token, we could quote Christoffel symbols of the cosmological metric as
\be
\left\{ {}^{\,\alpha}_{\mu\nu} \right\} \approx Nu^\alpha u_\mu u_\nu+ H u^\alpha h_{\mu\nu} - 2H h^\alpha{}_{(\mu}u_{\nu)}\,. 
\ee
The constraint (\ref{constraint}) imposed upon the affine connection (\ref{connection}) results in the 6 equations $R_i=0$ \cite{DAmbrosio:2021pnd} 
\be \label{constraints}
\dot{K}_5  =  0\,,\quad
\dot{K}_3  + K_3K_4 - K_1K_3  =  0\,, \quad
\dot{K}_2 + K_2K_4 -  K_1K_2  =  0\,, \quad
-k + K_2K_3 -K_5^2 =  0\,, \quad
K_3 K_5  =  0\,, \quad
K_4 K_5 =  0\,,
\ee
where $k$ is the spatial curvature parameter as used in this article. The first equation implies that $K_5$ is a constant $c$. If we set $c=0$, there remain 3 equations for 4 unknowns. If we further set $K_2=0$, in the flat case there remains 1 equation for 3 unknowns. Thus, we readily see that there at least exist more general parallel geometries than in the metric or in the symmetric special cases. A more thorough analysis of the 5 branches of connection solutions is  Ref.\cite{Heisenberg:2022mbo}. These are now understood as bifurcations of the 2 branches for the solutions for the fundamental matrix $\hat{\Lambda}$ we deduced in Section \ref{parallelism}.
We can check explicitly the forms of the functions in (\ref{ctorsion}) and in (\ref{cnonmetricity}),
\bs
\label{cqt}
\be
T_1  =  \frac{K_4-K_3}{n}\,, \quad T_2 = \frac{K_5}{a}\,, \quad
Q_1 = N-\frac{K_1}{n}\,, \quad 
Q_2 = \frac{K_4}{n} - H\,, \quad 
Q_3 = \frac{K_3}{n}-\frac{nK_2}{a^2}\,,
\ee
i.e.
\be
T_1  =  \hK_4-\hK_3\,, \quad T_2 = \hK_5\,, \quad
Q_1 = N-\hK_1\,, \quad 
Q_2 = \hK_4 - H\,, \quad 
Q_3 = \hK_3 - \hK_2\,,
\ee
\es
in terms of the 2 metric functions and the 5 connection functions in either convention. 
It is worth explicitly writing the five functions $K_i$ introduced in \cite{Hohmann:2020zre} in terms of the functions $\mu$ and $\sigma$ defining our cosmological frames. We can write the relation directly for the regular curved configuration
\be
{K}_1=\partial_0\log \mu,\quad {K}_2=-\frac{k\ell\sigma}{\mu},\quad {K}_3=\frac{\mu}{\ell\sigma},\quad {K}_4=\partial_0\log\sigma,
\ee
 while $K_5$ is only non-vanishing (and equal to the curvature parameter $-\sqrt{k}$) for the exceptional configuration. From these expressions we can again obtain the relation for the other branches by taking the appropriate limits. Written in terms of $\sigma$ and $\mu$, the equations (\ref{constraints}) become identities.

The G$_\parallel$R theory in the cosmological minisuperspace is then, equivalently with (\ref{isolag}), 
\be \label{isolag2}
L  =  -\frac{3m_P^2}{2}\lb 2H^2 - H \lp \hK_2 + 3\hK_3\rp - \lp N - \hK_1\rp \lp \hK_2-\hK_3\rp 
 + \hK_2\lp 2\hK_3-\hK_4\rp
+ \hK_3\hK_4 - 2\hK_5^2\rb\,. 
\ee
The constitutive laws are then read
\bs
\ba
S_\alpha{}^{\mu\nu} 
& = & -m_P^2\lp 2H-\hK_2-\hK_3\rp h_\alpha{}^{[\mu}u^{\nu]} -
\frac{1}{2}m_P^2 \hK_5\epsilon_\alpha{}^{\mu\nu}\,, \\
P^\alpha{}_{\mu\nu} & = & \frac{m_P^2}{4}\lb 3\lp \hK_3-\hK_2\rp u^\alpha u_\mu u_\nu + 
 \lp 4H - \hK_2 - 3\hK_3\rp u^\alpha h_{\mu\nu} 
 + 2\lp H + N - \hK_1-2\hK_3+\hK_4\rp h^\alpha{}_{(\mu}u_{\nu)}\rb\,. 
\ea
\es
The metrical and the canonical energy tensors are given by, respectively,
\bs
\ba
t^\mu{}_\nu
& = & -3m_P^2 \lp H^2 - 2H \hK_3 + \hK_3^2 + \hK_5^2\rp u^\mu u_\nu \nn \\ & - &
\lb 3H^2 - H\lp \hK_2 + 3\hK_3 + 2\hK_4\rp - \lp N - \hK_1\rp\lp \hK_2-\hK_3\rp 
+ \hK_2\hK_3 + 2\hK_3\hK_4 + \hK_5^2\rb h^\mu{}_\nu 
\,, \\ 
G^\mu{}_\nu 
& = &  t^\mu{}_\nu + m_P^2\lb H\lp 2\hK_3-2\hK_4\rp - \hK_2\lp \hK_3-\hK_4\rp 
- \hK_3^2 + \hK_3\hK_4 + 2\hK_5^2\rb h^\mu{}_\nu\,.
\ea
\es
Let us now have a look at the field equation. Define the short-hand notation for the superpotential $\mS$ and the axial torsion $\mA$ 
\be
S_\alpha{}^{\mu\nu} = \mS h_\alpha{}^{[\mu}u^{\nu]} + \mA \epsilon_\alpha{}^{\mu\nu}\,,
\quad \text{where} \quad \mS = m_P^2\lp 2 H - \hK_2 - \hK_3\rp\,, \quad 
\mA = -\frac{m_P^2}{2}\hK_5\,. 
\ee
The G$_\parallel$R field equation then takes the form
\be
T^\mu{}_\nu = 3\hK_3 \mS u^\mu u_\nu
- \lb {\mS}' + \lp 3H-\hK_4 \rp \mS + 8m_P^{-2}\mA^2 \rb h^\mu{}_\nu\,.
\ee
The connection equation $t^\mu{}_\nu=0$ is simplified when implementing the constraints (\ref{constraints}). This leads to the 5 branches of solutions arrived at in Ref.\cite{Heisenberg:2022mbo}: the flat Branches 1a, 1b, 1c and   
the curved Branches 2a, 2b. 
Below we consider the physical branch of solutions to (\ref{constraints}) in the cases $k=0$ (Branch 1c) and $k \neq 0$ (Branch 2b), respectively. The 2 non-canonical spatially flat branches of solutions are considered in the appendices \ref{branch1a} and \ref{branch1b}, and the non-canonical spatially flat branch is considered in appendix \ref{branch2a}. 

\section{Spatially flat canonical frame}

The physical branch of solutions is the flat $k=0$ Branch 1c wherein $K_2=K_5=0$, and the remaining 3 functions satisfy $\dot{{K}}_3/K_3 = K_1 - K_4$. We obtain from (\ref{cqt}) that
\be
Q_1 = -\hK_4 - (\log{\hK_3})'\,, \quad Q_2 = \hK_4-H\,, \quad Q_3 = \hK_3\,,
\quad T_1 = \hK_4-\hK_3\,, \quad T_2 = 0\,. 
\ee
The boundary term contribution to the Lagrangian is given by $K_3$,
\be
L = -3m_P^2H^2 + \frac{3 m_P^2}{2}\lp {\hK}'_3 + 3H \hK_3 \rp\,.
\ee
The constitutive laws are
\bs
\ba
S_\alpha{}^{\mu\nu} & = & m_P^2\lp 2H - \hK_3\rp h_\alpha{}^{[\mu}u^{\nu]}\,, \\
P^\alpha{}_{\mu\nu} & = & \frac{m_P^2}{4}\lb 3\hK_3u^\alpha u_\mu u_\nu - \lp 4H - 3\hK_3\rp u^\alpha h_{\mu\nu} + 2\lp H-{\hK}'_3/\hK_3 - 2\hK_3\rp h^\alpha{}_{(\mu}u_{\nu)}\rb\,,
\ea
\es
and the energy tensors are 
\bs
\ba
t^\mu{}_\nu & = & -3m_P^2\lp H -\hK_3\rp^2  u^\mu u_\nu - m_P^2\lp 3H^2 - {\hK}_3'
-3H\hK_3 - 2H\hK_4 + \hK_3\hK_4\rp h^\mu{}_\nu\,,
\\
G^\mu{}_\nu & = & -3m_P^2 \lp H -\hK_3\rp^2  u^\mu u_\nu - m_P^2\lp 3H^2 - {\hK}_3'-5H\hK_3+\hK_3^2\rp h^\mu{}_\nu\,.   
\ea
\es
This solution can be smoothly transformed into a canonical frame from
a General Linear frame parameterised by 2 functions of time. In a canonical frame, these functions are fixed into the Diff-invariant expressions $\hK_3=H$, $\hK_4=-(\log{H})'$. It follows that $\hK_1=N$. 

\section{Spatially curved canonical frame} 

When $k \neq 0$, the functions in the physical Branch 1b are generalised to 
\be
Q_1 = -\hK_4 - {\hK}'_3/\hK_3\,, \quad Q_2 = \hK_4-H\,, \quad Q_3 = \hK_3 - \frac{k^2}{a^2 \hK_3}\,,
\quad T_1 = \hK_4-\hK_3\,, \quad T_2 = 0\,. 
\ee
Again, $\hK_4$ is decoupled from the Lagrangian,
\be
L = -3m_P^2\lp H^2 - \frac{k}{a^2}\rp  + \frac{3 m_P^2}{2}\lp {\hK}'_3 + 3H \hK_3 \rp + \frac{3 m_P^2 k}{2a^2}\lp \frac{{\hK}'_3}{\hK_3^2} - \frac{H}{\hK_3} \rp \,.
\ee
The constitutive laws are
\bs
\ba
S_\alpha{}^{\mu\nu} & = & m_P^2\lp 2H + \frac{k}{a^2\hK_3} - \hK_3\rp h_\alpha{}^{[\mu}u^{\nu]}\,, \\
P^\alpha{}_{\mu\nu} & = & \frac{m_P^2}{4}\lb 3\lp \hK_3 + \frac{k}{a^2\hK_3}\rp u^\alpha u_\mu u_\nu - \lp 4H - 3\hK_3 - \frac{k}{a^2\hK_3}\rp u^\alpha h_{\mu\nu} + 2\lp H-{\hK}'_3/\hK_3 - 2\hK_3\rp h^\alpha{}_{(\mu}u_{\nu)}\rb\,,
\ea
\es
and the energy tensors are 
\bs
\ba
t^\mu{}_\nu & = & -3m_P^2\lp H -\hK_3\rp^2  u^\mu u_\nu - m_P^2\lp 3H^2 -{\hK}'_3
-3H\hK_3 - 2H\hK_4 + \hK_3\hK_4\rp h^\mu{}_\nu \nn \\
& + & \frac{m_P^2 k}{a^2\hK_3}\lp H - \hK_3'/\hK_3 - \hK_3 - \hK_4\rp h^\mu{}_\nu\,, \label{curvedt}
\\
G^\mu{}_\nu & = & -3m_P^2 \lp H -\hK_3\rp^2  u^\mu u_\nu - m_P^2\lp 3H^2 - {\hK}'_3-5H\hK_3+\hK_3^2\rp h^\mu{}_\nu \nn \\
& + & \frac{m_P^2 k}{a^2\hK_3}\lp H - \hK_3'/\hK_3 - 2\hK_3\rp h^\mu{}_\nu\,.   
\ea
\es
The spatial curvature contributes a new term to the reference frame energy, the \2 line of (\ref{curvedt}). The connection in a canonical frame is otherwise the same as in the case $k=0$, with the nonzero functions in (\ref{connection}) $\hK_1=N$, $\hK_3=H$, and $\hK_4=-(\log{H})'$, 
but now also the $K_2=k/H$ is forced to be nonzero. This choice consistently also vanishes the new term in (\ref{curvedt}). A zero-energy frame is obtained by setting $\hK_3 = H \pm \sqrt{H^2 + k/a^2}$.

\section{The coincident frame is non-canonical}
\label{coincident}

If $K_i=0$, we obtain the constitutive laws
\bs
\ba
S_\alpha{}^{\mu\nu}  & = & 2\mpl^2 H h_\alpha{}^{[\mu}u^{\nu]}\,, \\
P^\alpha{}_{\mu\nu} & = & -\mpl^2 H u^\alpha h_{\mu\nu}
-\frac{1}{2}\mpl^2\lp H + N \rp h^\alpha{}_{(\mu}u_{\nu)}\,,
\ea
\es
and the metric energy tensor is the canonical one,
\be
G^\mu{}_\nu = t^\mu{}_\nu = -3\mpl^2 H^2\lp u^\mu u_\nu + h^\mu{}_\nu\rp\,.  
\ee
In this frame, there is an effective stiff fluid whose energy exactly cancels the material energy, $ u_\mu u_\nu  (T^{\mu\nu} + t^{\mu\nu})  =0$. The pressure of matter is not canceled $h_{\mu\nu}(T^{\mu\nu} + t^{\mu\nu}) \neq 0$ is but sources the gravitational field in this frame. In appendix \ref{mtegr} we show that in the metric teleparallel equivalent of GR, the inertial effective source behaves like radiation fluid instead of a stiff fluid.

In the coincident frame, the quantities above reduce to the historical pseudotensor approach. They are called pseudotensors, since they transform non-covariantly. We refer to this solution as rather the coincident frame (vs coincident gauge), because it is a non-canonical frame and cannot be reached from a physical solution by a gauge transformation.
N. Rosen had noted that in these coordinates the total energy of the universe is zero  \cite{rosen} i.e. $u_\mu u_\nu  (T^{\mu\nu} + t^{\mu\nu})=0$, and the similar result has also been noticed in the metric teleparallel formulation, e.g. \cite{vargas,deLaRica:2009kk,Ulhoa:2010wv}. A more extensive study of the latter is found in Ref.\cite{Nester:2008xd}.

\section{The absence of canonical frame in symmetric teleparallelism}
\label{stgr}
In this appendix we will more details on the (non-)existence of a canonical frame in the symmetric teleparallel case, which is defined by the additional condition of vanishing torsion. Thus, we need to impose $T_1=T_2=0$. The requirement $T_2=0$ again excludes the exceptional curved configuration, while the other configurations remain oblivious to it. Regarding the condition $T_1=0$, only the non-trivial flat configuration I and the regular curved configuration can satisfy it.

The nonmetricity tensor takes the form
\be
Q_{\alpha\mu\nu} = 2Q_1u_\alpha u_\mu u_\nu + 2Q_2 u_\alpha h_{\mu\nu} + 2 Q_3 h_{\alpha(\mu}u_{\nu)}\,. 
\ee
where the $Q_a$ can only depend on the time coordinate.
We obtain the traces
\bs
\ba
Q_\alpha & = & \lp -2Q_1 + 6 Q_2\rp u_\alpha\,, \\
\tilde{Q}_\alpha & = & \lp -2 Q_1 + 3Q_3\rp u_\alpha\,.
\ea
\es
Using this Ansatz, we can compute the non-metricity contributions in the explicit expression (\ref{einertial}),
\bs
\label{cterms}
\ba
C_1 & = & Q_{\alpha\mu\nu}Q^{\alpha\mu\nu} = -4Q_1^2 - 12 Q_2^2 - 6Q_3^2\,, \\
Q_{\nu\alpha\beta}Q^{\mu\alpha\beta} & = &  4\lp Q_1^2 + 3Q_2^2\rp u^\mu u_\nu - 2Q_3^2 h^\mu{}_\nu\,, \\
C_2 & = & Q_{\nu\alpha\beta}Q^{\alpha\beta\mu} = -4Q_1^2 - 12Q_2Q_3 - 3Q_3^2\,, \\
Q_{\nu\alpha\beta}Q^{\alpha\beta\mu} & = & \lp 4Q_1^2+6Q_2Q_3\rp u^\mu u_\nu - Q_3\lp 2 Q_2+Q_3\rp h^\mu{}_\nu\,, \\
C_3 & = & Q_\alpha Q^\alpha = -4\lp Q_1- 3Q_2\rp^2\,, \\
Q_{\nu\alpha\beta}Q^{\mu}g^{\alpha\beta} & = & 4\lp Q_1 - 3Q_2\rp^2 u^\mu u_\nu\,, \\
C_4 & = & \tilde{Q}_\alpha\tilde{Q}^\alpha = -\lp 2Q_1-3Q_3\rp^2\,, \\
\tilde{Q}^\alpha Q_\nu{}^\mu{}_\alpha & = & 2Q_1\lp  2Q_1-3Q_3\rp u^\mu u_\nu + \lp 2 Q_1-3Q_3\rp Q_3 h^\mu{}_\nu\,, \\
C_5 & = & Q_\alpha \tilde{Q}^\alpha = -2Q_1\lp 2Q_1-6Q_2-3Q_3\rp  - 18 Q_2Q_3\,, \\
\lp Q^\alpha Q_\nu{}^\mu{}_\alpha + Q_\nu\tilde{Q}^\mu\rp & = & 2\lb Q_1\lp 4Q_1-12Q_2-3Q_3\rp + 9 Q_2Q_3\rb u^\mu u_\nu + 2\lp Q_1 - 3 Q_2\rp Q_3 h^\mu{}_\nu\,. 
\ea
\es
Then, we'll show that there is no canonical frame for the universe in symmetric teleparallelism, exploiting Hohmann's derivations \cite{Hohmann:2021ast}. From (\ref{cterms}) we can compute
\be
Q = -\frac{1}{4}\lp A  - C\rp + \frac{1}{2}\lp B - E\rp = -3\lp 2Q_2^2 + Q_1Q_3 - Q_2 Q_3\rp\,,
\ee
and 
\be
Q_{\nu\alpha\beta}P^{\mu\alpha\beta} = 3\lp 2Q_2^2 + \frac{1}{2}Q_1Q_3- \frac{1}{2}Q_2Q_3\rp u^\mu u_\nu  - \frac{1}{2}\lp Q_1-Q_2\rp Q_3 h^\mu_\nu\,.
\ee
From these we obtain that
\be
t^\mu{}_\nu  = -3Q_2^2u^\mu u_\nu - \lp 3Q_2^2 + Q_1Q_3 - Q_2Q_3\rp h^\mu_\nu\,.
\ee
Thus, the frame is canonical iff $Q_2=0 \land (Q_1=0 \lor Q_3=0)$. Hohmann reported 3 branches of solutions in symmetric teleparallelism, page 6 in \cite{Hohmann:2021ast}. 
\begin{itemize}
\item In the \1 branch, the above implies that either $\dot{H} =-H^2$ or $H^2=-k/a^2$. Both imply the curvature-dominated universe with $w=-1/3$. 
\item In the \2 branch, we solve the different equations but find the same implications.
\item The \3 branch is the ``rather cumbersome'' one which in the flat $k=0$ case reduces to solution with $Q_2=-H$, so that any expansion is non-canonical.
\end{itemize}
Thus, symmetric teleparallelism is too restrictive. 

\section{The absence of canonical frame in metric teleparallelism}
\label{mtegr}

The torsion terms contributing to the canonical energy tensor can be computed as follows. (We use the shorthand notation: $A_i^{\alpha\beta\mu}$ is the contribution from the 
term $A_i$ in the Lagrangian to the tensor $S^{\alpha\beta\mu}$). 
\bs
\label{tcosmo}
\ba
A_1 & = & T_{\alpha\mu\nu}T^{\alpha\mu\nu} = -6T_1^2 + 24T_2^2\,, \\
T_{\alpha\beta\nu}A_1^{\alpha\beta\mu} & = &  6T_1^2 u^\mu u_\nu - 2\lp T_1^2 -8T_2^2\rp h^\mu{}_\nu\,, \\
A_2 & = & T_{\alpha\mu\nu}T^{\mu\alpha\nu} = -3\lp T_1^2+8T_2^2\rp\,, \\
T_{\alpha\beta\nu}A_2^{\alpha\beta\mu} & = & 3T_1^2 u^\mu u_\nu - \lp T_1^2 + 16T_2^2\rp h^\mu{}_\nu\,, \\
A_3 & = & T_\alpha T^\alpha = -9T_1^2\,, \\
T_{\alpha\beta\nu}A_3^{\alpha\beta\mu} & = & 3T_1^2\lp 3u^\mu u_\nu - h^\mu{}_\nu\rp\,. 
\ea
\es
From (\ref{tcosmo}) we can compute, in metric teleparallelism
\bs
\be
L = -\frac{1}{4}A_1 - \frac{1}{2}A_2 + A_3 = 
-6\lp T_1^2 - T_2^2\rp\,,
\ee
and 
\be
T_{\alpha\beta\nu}S^{\alpha\beta\mu} = 6T_1^2 u^\mu u_\nu - 2\lp T_1^2-2T_2^2\rp h^\mu{}_\nu\,.
\ee
\es
Thus, we obtain 
\be
t^\mu{}_\nu = -2\lp T_1^2 + T_2^2\rp\lp 3u^\mu u_\nu + h^\mu{}_\nu\rp\,.
\ee
In metric teleparallel cosmology, we have $T_1=-H/n$, $T_2=0$, and thus an expanding universe can only be described in a non-canonical frame. The tensor $t^\mu{}_\nu$ is traceless, corresponding to inertial energy in the form of radiation. This has been noted previously \cite{deLaRica:2009kk,Formiga:2021inl}. 

There is a history of attempts to define the gravitational energy in the metric teleparallel reformulation of GR \cite{Moller:233632}, see e.g. \cite{deAndrade:2000kr,deLaRica:2009kk,Obukhov:2006sk,Lucas:2009nq,Krssak:2015rqa,Krssak:2015lba,Formiga:2020wsk,Bahamonde:2021gfp,Formiga:2021inl,Maluf:2023rwe}. Some of the considerations on the metric-teleparallel spin connection conform with the concept of canonical frame according to G$_\parallel$R, though may be insufficient for the {\it local} resolution of the physical frame. Notably, the “holographic renormalisation" -type prescriptions have been used to regularise metric-teleparallel gravity actions and shown to give reasonable results for charges considered as global volume integrals  \cite{Obukhov:2006sk,Lucas:2009nq,Krssak:2015rqa,Krssak:2015lba,Krssak:2023nrw}. 


\section{The non-canonical Branch 1a}
\label{branch1a}

This is the flat $k=0$ branch wherein the only nonzero $K_i$ are $K_1$ and $K_4$. We redefine the variables such that the 2 functions $K_1$ and $K_4$ are traded for $X=Q_2$ and $Y=Q_1$. Then, from
(\ref{potentials}), this branch is specified by the functions
\be
Q_1= Y\,, \quad Q_2 = X\,, \quad Q_3 =0\,, \quad T_1 = H + X\,, \quad T_2 = 0\,. 
\ee
The Lagrangian (\ref{isolag}) depends neither on $X$ nor $Y$, but $L = -6H^2$.
The excitation tensor and the canonical energy tensor become
\bs
\ba
S_\alpha{}^{\mu\nu} & = & 2\mpl^2H h_\alpha{}^{[\mu}u^{\nu]}\,, \\
P^\alpha{}_{\mu\nu} & = & -\mpl^2H u^\alpha h_{\mu\nu} + \frac{1}{2}\mpl^2 \lp X + Y\rp h^\alpha{}_{(\mu\nu)} \nn \\
& = & -\mpl^2H u^\alpha h_{\mu\nu} - \frac{1}{2}\mpl^2\lp H + N - \hK_1 + \hK_4\rp\,. 
\ea
\es
In this branch, the superpotential is frame-independent. The energy tensors are
\bs
\ba
t^\mu{}_\nu & = & -3\mpl^2 H^2 u^\mu u_\nu - \mpl^2 H\lp H - 2X\rp h^\mu{}_\nu \nn \\
& = & -3\mpl^2 H^2 u^\mu u_\nu - \mpl^2 H\lp 3H-2\hK_4\rp h^\mu{}_\nu\,, \\
G^\mu{}_\nu & = & -3\mpl^2 H^2\lp u^\mu u_\nu + h^\mu{}_\nu\rp\,.
\ea
\es
In this branch, the energy of matter is cancelled by the negative inertial energy effectively carried by the geometric fields. 

\section{The non-canonical Branch 1b}
\label{branch1b}

This is the flat $k=0$ branch wherein $K_3=K_5=0$, and the remaining 3 functions satisfy $\dot{{K}}_2/K_2 = K_4 - K_1$. 
If we now define $K_4=n(X+H)$ and $K_2=-a^2Y/n=-\hK_2$, we obtain from (\ref{potentials}) that
\be
Q_1 = H-X + (\log{Y})'\,, \quad Q_2 = X\,, \quad Q_3 = Y\,, \quad T_1 = H + X\,, \quad T_2 = 0\,.  
\ee
Now the Lagrangian (\ref{isolag}) depends on $Y$ as
\be \label{lag1b}
L = -3\mpl^2\lp H^2 + \frac{3}{2}HY + \frac{1}{2}{Y}'\rp\,.
\ee
The EoM for $Y$ derived from this is consistently trivial. The excitation tensor and 
\bs
\ba
S_\alpha{}^{\mu\nu} & = & \mpl^2 \lp 2H + Y\rp h_\alpha{}^{[\mu}u^{\nu]}\,, \\
P^\alpha{}_{\mu\nu} & = & \frac{\mpl^2}{4}\lb 3Y^2 u^\alpha u_\mu u_\nu-\lp 4H+Y\rp u^\alpha h_{\mu\nu} + 2 \lp Y' + 3HY\rp h^\alpha{}_{(\mu}u_{\nu)}\rb\,. 
\ea
\es
The two energy tensors are now
\bs
\ba
t^\mu{}_\nu & = & -3\mpl^2 H^2 u^\mu u_\nu - \mpl^2\lb H\lp H-2X+2Y\rp + {Y}' - XY \rb h^\mu{}_\nu\,, \\
G^\mu{}_\nu & = &
-3\mpl^2 H^2 u^\mu u_\nu - \mpl^2\lb 3H\lp H + Y\rp + {Y}'\rb h^\mu{}_\nu\,.  
\ea
\es
Again the metric energy is minus the material energy. 

\section{The non-canonical Branch 2a}
\label{branch2a}

This is the spatially curved $k \neq 0$ branch wherein $K_2=K_3=0$ and we can set $K_5=\sqrt{-k}$. The 5 functions  (\ref{potentials}) are then as follows.
\be
Q_1 = N-\hK_1\,, \quad Q_2 = \hK_4-H\,, \quad Q_3 = 0\,, \quad T_1 = \hK_4\,, \quad T_2 = \sqrt{-k}/a\,. 
\ee
The Lagrangian of G$_\parallel$R $L=-3\mpl^2(H + \sqrt{-k}/a)(H - \sqrt{-k}/a)$. 
The branch 2a is the generalisation of branch 1a with spatial curvature $k \neq 0$.  
The constitutive laws are 
\bs
\ba
S_\alpha{}^{\mu\nu} & = & 2\mpl^2 H h_\alpha{}^{[\mu}u^{\nu]} - \frac{\mpl^2}{2}(\sqrt{-k}/a)\epsilon_\alpha{}^{\mu\nu}\,, \\
P^\alpha{}_{\mu\nu} & = & -\mpl^2H u^\alpha h_{\mu\nu} - \frac{\mpl^2}{2}\lp H + N - \hK_1 + \hK_4\rp h^\alpha{}_{(\mu}u_{\nu)}\,,
\ea
\es
and the energy tensors are
\bs
\ba
t^\mu{}_\nu & = &  -3\mpl^2\lb  H^2 - \frac{k}{a^2}\rb u^\mu u_\nu - \mpl^2 \lb 3H^2 - \frac{k}{a^2} - 2H\hK_4\rb h^\mu{}_\nu\,, \\
G^\mu{}_\nu & = & -3\mpl^2\lb  H^2 - \frac{k}{a^2}\rb u^\mu u_\nu - \mpl^2 \lb 3H^2 + \frac{k}{a^2}\rb h^\mu{}_\nu\,.
\ea
\es
This family of solutions cannot be put into a canonical frame. The energy-free frame is obtained by setting $K_4=-(\log{H})'$, which leaves $K_1$ arbitrary.  

\section{Remarks on literature concerning energy (pseudo)tensors}
\label{literature}

In Section \ref{energytensors} we introduced the energy tensors $T^\mu{}_\nu$, $G^\mu{}_\nu$ and $t^\mu{}_\nu$, justifying their physical interpretation by the \1 fundamental law of interactions. 
In particular, $T^\mu{}_\nu$ describes material energy, and $t^\mu{}_\nu$ describes inertial energy. In the
standard metric-geometrical formulation of GR, the field equation is not expressed in the canonical form, and therefore the identification of the observables
is much less straightforward. However, it should be recognised that our interpretation nevertheless agrees with the well-established 
understanding of energy in the context of GR. We could not communicate this understanding more clearly and convincingly than the classic textbook 
\cite{Misner:1973prb}, and shall therefore quote two paragraphs from their §20.4, titled {\it Why the energy of the gravitational field cannot be localized}:  
\newline
\newline
{\it To ask for the amount of electromagnetic energy and momentum in an element of a 3-volume makes sense. First, there is one and only
one formula for this quantity. Second, and more important, this energy-momentum in principle ``has weight''. It curves space. It serves as source
term on the righthand side of Einstein's field equations. It produces relative geodesic deviation of two nearby world lines that pass through
the region of space in question. It is observable.}
\newline
\newline
{\it Not one of these properties
does ``local gravitational energy-momentum'' possess. There is no unique formula for it, but a multitude of quite distinct formulas. [...] 
``local gravitational energy-momentum'' has no weight. It does not curve space. It does not serve as a source term on the righthand side of 
the Einstein's field equations. It does not produce any relative geodesic deviation of two nearby world lines that pass through the region of space
in question. It is not observable.} 
\newline
\newline
In our formulation, the localisable energy-momentum which {\it makes sense} is described by $T^\mu{}_\nu$. The ``local gravitational energy-momentum'' that is 
sandwiched between quotation marks because it {\it has no weight} and  {\it is not observable} is in our formulation described by the tensor $t^\mu{}_\nu$. Whilst in the context of GR it has indeed been described by {\it a multitude of quite distinct formulas} in terms of pseudotensors, in G$_\parallel$R those are understood to represent the same tensor in different frames, as was explicitly shown to be the case for the formulas of Bergmann-Thomson, Einstein, von Freud, Landau-Lifshitz, Maluf, Papapetrou, and Weinberg \cite{Gomes:2022vrc}.

An achievement of G$_\parallel$R is the elimination of this unobservable from the theory of GR. This demystification of gravity undermines the dogma that the absence of well-defined charges in GR reflects, not an incompleteness of the formulation of the theory, but the fundamental property of Nature. This view, called ``the geometrical dogma'' \cite{Koivisto:2022nar}, and echoed in the recent pamphlet \cite{Golovnev:2023ogu}, seems incompatible with Einstein's view of gravity and with the idea of gravity as a gauge theory of translations. In both form and in spirit, the parallel theory of gravity is more akin to the original theory of GR \cite{https://doi.org/10.1002/andp.19163540702} than the modified, mathematico-geometrical $f(\mathcal{R})=\mathcal{R}$ model usually taken for ``the GR''. However, a key difference introduced in G$_\parallel$R wrt the original theory is the disambiguation of the reference frame and the coordinate system. 

Another difference wrt early attempts at definitions of energy is the realisation of the quasilocal property of the physical charges \cite{Szabados:2004xxa}. This is also a crucial distinction of the G$_\parallel$R with respect to many metric-teleparallel energy prescriptions, e.g. \cite{Moller:233632,deAndrade:2000kr,Nashed:2006yw,Sousa:2008av,So:2008kr,deLaRica:2009kk,Obukhov:2006sk,Lucas:2009nq,Nashed:2010ocg,Sharif:2010zz,Nashed:2011fg,Nashed:2012zz,Nashed:2013eva,Krssak:2015rqa,Krssak:2015lba,Mourad:2019cza,Formiga:2020wsk,Bahamonde:2021gfp,Formiga:2021inl,Aygun:2023why,Maluf:2023rwe,Krssak:2023nrw,Fiorini:2023axr}. It might be that some physical counterpart could be established to the energy-momentum densities described by the (pseudo)tensors used in such prescriptions, perhaps related to the gas of particles due to Unruh radiation seen by accelerated observers. Presently this is mere speculation, and we may only remind of the fact as it was stated now 50 years ago 1973 \cite{Misner:1973prb}:
\newline
\newline
{\it Anybody who looks for a magic formula for the ``local gravitational energy-momentum'' is looking for the right answer to the wrong question. Unhappily, enormous time and effort were devoted in the past to trying to ``answer this question'' before investigators realized the futility of the enterprise.}

\bibliography{TeleCosmo}

\end{document}